\documentclass[numberedappendix]{aastex}
\usepackage{emulateapj5}
\usepackage{graphicx}

\shortauthors{Hopkins et al.}
\shorttitle{}

\received{2003 June 27}
\begin{document}

\title{Star formation rate indicators in the Sloan Digital Sky Survey}

\author{A. M. Hopkins\altaffilmark{1,2}, C. J. Miller\altaffilmark{3},
        R. C. Nichol\altaffilmark{3}, A. J. Connolly\altaffilmark{1},
        M. Bernardi\altaffilmark{1,3}, P. L. G{\'o}mez\altaffilmark{3},
        T. Goto\altaffilmark{4}, C. A. Tremonti\altaffilmark{5},
        J. Brinkmann\altaffilmark{6}, {\v Z}. Ivezi{\'c}\altaffilmark{7},
        D. Q. Lamb\altaffilmark{8}
}

\affil{
\begin{enumerate}
\item Dept.\ of Physics and Astronomy, University
 of Pittsburgh, 3941 O'Hara Street, Pittsburgh, PA 15260
\item Hubble Fellow; email ahopkins@phyast.pitt.edu
\item Dept.\ of Physics, Carnegie Mellon University,
  5000 Forbes Avenue, Pittsburgh, PA 15213
\item Institute for Cosmic Ray Research, University of Tokyo,
  Kashiwanoha, Kashiwa, Chiba 277-0882, Japan
\item Dept.\ of Physics and Astronomy, Johns Hopkins University,
  3400 North Charles Street, Baltimore, MD 21218
\item Apache Point Observatory, 2001 Apache Point Road,
  P.\ O.\ Box 59, Sunspot, NM 88349-0059
\item Princeton University, Department of Astrophysical Sciences,
  Princeton, NJ 08544-1001
\item Dept.\ of Astronomy and Astrophysics,
  University of Chicago, 5640 S.\ Ellis Ave, Chicago, IL 60637
\end{enumerate}
}

\begin{abstract}
The Sloan Digital Sky Survey (SDSS) first data release provides a database
of $\approx 106000$ unique galaxies in the main galaxy sample with
measured spectra. A sample of star-forming (SF) galaxies are identified
from among the 3079 of these having 1.4\,GHz luminosities from FIRST,
by using optical spectral diagnostics. Using 1.4\,GHz luminosities as a
reference star formation rate (SFR) estimator insensitive to obscuration
effects, the SFRs derived from the measured SDSS H$\alpha$, [O{\sc ii}]
and $u$-band luminosities, as well as far-infrared luminosities from
IRAS, are compared. It is established that straightforward corrections for
obscuration and aperture effects reliably bring the SDSS emission line and
photometric SFR estimates into agreement with those at 1.4\,GHz, although
considerable scatter ($\approx 60\%$) remains in the relations. It thus
appears feasible to perform detailed investigations of star formation
for large and varied samples of SF galaxies through the available
spectroscopic and photometric measurements from the SDSS. We provide
herein exact prescriptions for determining the SFR for SDSS galaxies. The
expected strong correlation between [O{\sc ii}] and H$\alpha$ line fluxes
for SF galaxies is seen, but with a median line flux ratio $F_{\rm
[OII]}/F_{\rm H\alpha}=0.23$, about a factor of two smaller than that
found in the sample of \citet{Ken:92}. This correlation, used in deriving
the [O{\sc ii}] SFRs, is consistent with the luminosity-dependent relation
found by \citet{Jan:01}. The median obscuration for the SDSS SF systems
is found to be $A_{\rm H\alpha}=1.2\,$mag, while for the radio detected
sample the median obscuration is notably higher, 1.6\,mag, and with a
broader distribution.
\end{abstract}


\section{Introduction}
\label{int}

The current star formation rate (SFR) of a galaxy is one of many
important parameters used in developing our understanding of galaxy
evolution. Various different indicators of galaxy SFR exist at
different wavelengths, and include H$\alpha$ and [O{\sc ii}] emission
line luminosities, ultraviolet continuum luminosity, far-infrared (FIR)
luminosity and radio luminosities \citep[see reviews by][]{Ken:98,Con:92}.
There are also strong suggestions that X-ray luminosity is an important
SFR indicator \citep{Gri:90,WG:98,GW:01,Ptak:01,Bra:01,Age:03}, although
very faint X-ray observations are typically required to detect X-ray
emission from star formation processes. In recent years several comparisons
between different star formation indicators at
varying wavelengths have been made, primarily establishing broad agreement
between each but with detailed discrepancies and large scatter
in the relations \citep{Bell:03,Buat:02,Hop:01,Sul:01,Sul:00,Cram:98}.
One of the limiting factors in these comparisons to date is the mostly
heterogeneous nature of the data being compared, and the relatively small
numbers of objects investigated, no more than a few hundred.

The Sloan Digital Sky Survey
\citep[SDSS,][]{Fuk:96,Gun:98,York:00,Hog:01,Sto:02,Smi:02,Pie:03} eliminates
these limitations. The SDSS is supplying the astronomical community with
images and spectra providing an immense resource for use in numerous studies
of galaxy, quasar, stellar and solar system properties. By identifying
star-forming (SF) galaxies catalogued by the SDSS, a very large, homogeneous
sample can be used to investigate the properties of star formation in
galaxies. To support such studies, we investigate herein the consistency of
SFRs derived from SDSS H$\alpha$ and [O{\sc ii}] line measurements with
those derived from the obscuration independent 1.4\,GHz and FIR luminosities.
Further, we empirically derive a non-linear calibration of SFR from the
SDSS $u$-band luminosity, which gives SFR estimates highly consistent with
the other four estimators. X-ray luminosities are not pursued here as an
SFR estimator since not enough deep X-ray data is presently available for an
exploration of a homogeneous X-ray detected SF galaxy sample.
Results from an analysis of ROSAT All Sky Survey (RASS) identifications with
SDSS galaxies in the early data release \citep{RSL:03} confirm that almost
all the RASS X-ray sources are identified with known galaxy clusters, quasars
or active galactic nuclei (AGN).

In \S\,\ref{sample} we describe the details of the current sample.
The five SFR indicators are each explored in \S\,\ref{sfr}, beginning
with the 1.4\,GHz estimator in \S\,\ref{sfrradio}.
The H$\alpha$ SFRs, including the obscuration and aperture corrections
required, are presented in \S\,\ref{hasfr}. SFRs from [O{\sc ii}], FIR
and $u$-band luminosities are each given in \S\S\,\ref{sectiono2},
\ref{sectionfir}, and \ref{sectionsfru} respectively. A discussion of
the absolute SFR calibrations and the properties of the radio detected
SF galaxies are explored in \S\,\ref{disc}. We summarise our
results in \S\,\ref{summ}. Throughout this paper we assume a
($\Omega_M=0.3,\Omega_\Lambda=0.7,H_0=70$) cosmology.

\section{Sample selection}
\label{sample}

The SDSS sample of spectroscopically observed main galaxies \citep{Str:02}
taken from the first data release \citep[DR1,][]{Aba:03} was used as the
starting point for constructing our sample. This catalogue of $\approx 106000$
unique galaxies yields 3079 galaxies with 1.4\,GHz measurements from the Faint
Images of the Radio Sky at Twenty centimeters catalogue
\citep[FIRST,][]{Whi:97}. The radio identification from FIRST is based on
positional matching within a radius of $1\farcs5$ of the SDSS position
\citep{Ive:02}. These galaxies form our primary sample for the investigations
presented here.

\begin{figure}
\centerline{\rotatebox{-90}{\includegraphics[width=12cm]{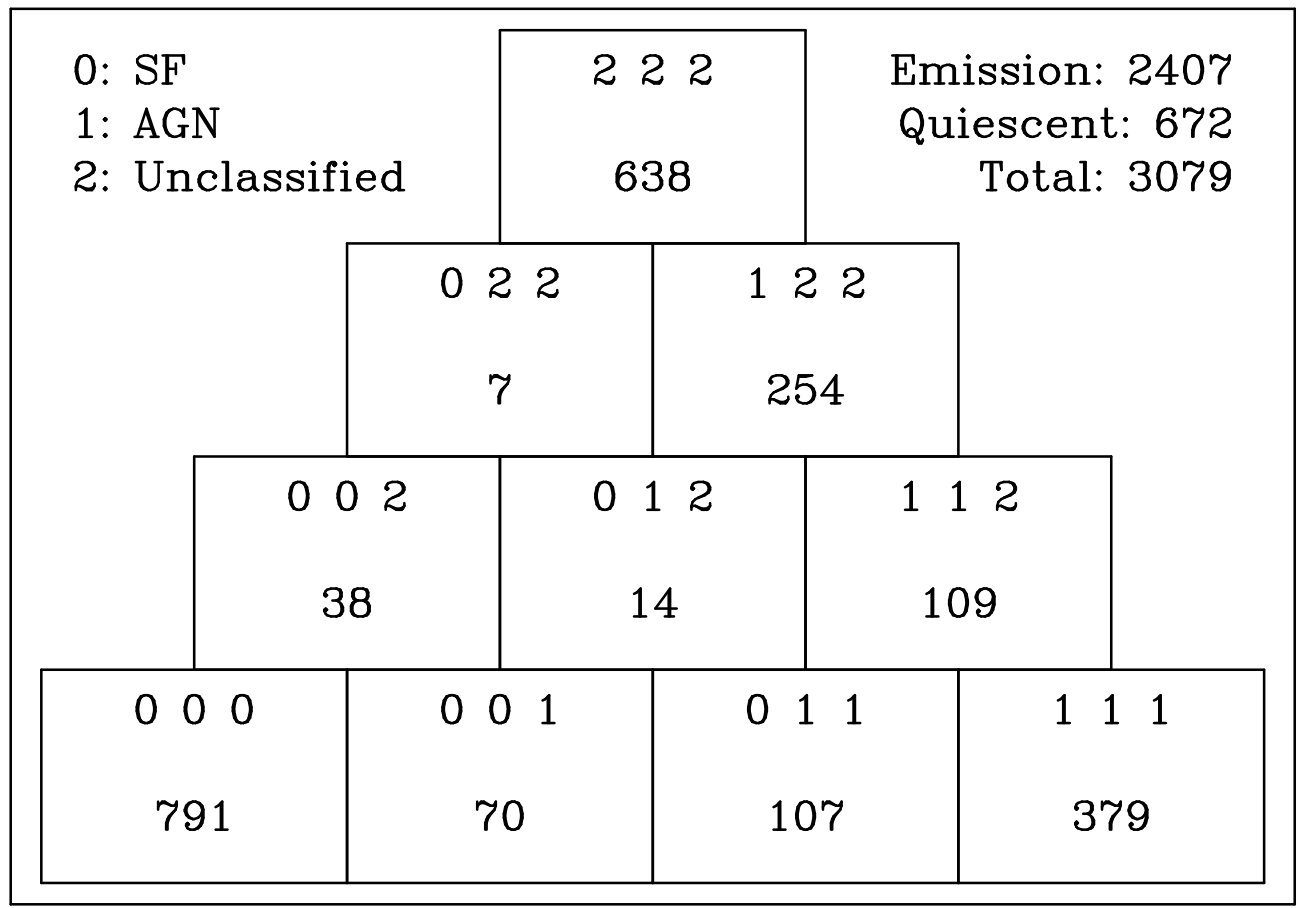}}}
\caption{Distribution of the spectroscopic diagnostic classifications for all
3079 DR1 main galaxies with 1.4\,GHz FIRST detections. The three numbers at
the top of each box are the flags (0, 1, or 2) indicating the classification
in the three spectral diagnostic diagrams. Below these is the number of
sources with each particular combination of flags (for example, there
are 791 sources with 0 for each flag).
There are 672 ``quiescent" systems, with none of the necessary emission
lines for classification.
 \label{pyramid}}
\end{figure}

Within our primary sample, we made use of three spectral diagnostic
diagrams to classify galaxies as either SF, active galactic nuclei (AGN)
or unclassified. Following \citet{Kew:01} and \citet{Mil:03},
the diagrams used were [O{\sc iii}]/H$\beta$ vs [N{\sc ii}]/H$\alpha$,
[O{\sc iii}]/H$\beta$ vs [S{\sc ii}]/H$\alpha$, and [O{\sc iii}]/H$\beta$ vs
[O{\sc i}]/H$\alpha$. We used the conservative requirements
that for a galaxy to be classified as SF or AGN it had to be so
classified in all three of the diagnostic diagrams, and remaining
unclassified otherwise. Only emission lines where the measured flux
was greater than twice the flux uncertainty (similar to a $2\sigma$
threshold) were considered in producing these classifications
\citep{Mil:03}. Additionally, some objects could also be classified as
AGN on the basis of having a large ratio of two emission lines for a
given diagram, even though the other pair of lines for that diagram were
not measurable \citep[see discussion in][]{Mil:03}. Saturated emission
lines will not affect our sample since saturated line parameters are
not measured, so systems with saturated H$\alpha$, for example, can not
be classified as SF galaxies. These systems are also extremely rare,
and mostly dominated by nuclear emission from very nearby galaxies,
so very few SF galaxies are likely to be omitted by excluding such
objects. Our primary sample consists of 791 SF galaxies, 379 AGNs, 672
``quiescent" galaxies (with no measurable diagnostic emission lines),
with the remaining 1237 galaxies either inconsistently classified,
or lacking the required emission lines to be able to be classified in
all three diagrams (see Figure~\ref{pyramid}). The ratio of AGN to SF
systems is relatively high as a result of requiring radio detection (and
hence a bias towards AGN systems) for inclusion in the sample. While a
number of the ``quiescent" galaxies are likely to be absorption line
systems, typical of old stellar populations in elliptical galaxies,
many are likely to be higher redshift objects whose emission lines fall
below the sensitivity limits of the SDSS.

Although true absorption line systems, particular those at higher redshifts
($z\gtrsim0.3$), will incorporate the luminous red galaxies
\citep[LRGs,][]{Eis:01}, our sample has been defined to exclude these objects,
being drawn from the main galaxy sample only \citep{Str:02}. This point is
relevant since LRGs were noted by \citet{Ive:02} to have a relatively high
fraction of radio counterparts. However, even if the LRGs are not excluded
a priori, the results presented here do not change at all, since almost none
of the LRGs satisfy our spectroscopic classification to be considered as SF
galaxies.

Of the systems with the necessary emission lines, relatively few show
conflicting classifications, as seen in Figure~\ref{pyramid}. The 177 galaxies
inconsistently classified between AGN and SF may be composite systems hosting
both types of activity \citep{Hill:01}, and are thus interesting in their
own right, although they are not further investigated here.

All the classified SF and AGN galaxies have measurable H$\alpha$ luminosity
($L_{\rm H\alpha}$) and Balmer decrements, a result of requiring H$\alpha$
and H$\beta$ for classification, and we restrict the current investigation to
the 791 galaxies classified as SF in all three spectral diagnostic
diagrams. To calculate $u$-band luminosities we chose
to use the Petrosian $u$-band absolute magnitudes, $M_u$,
(calculated using the k-corrections of \citep{Bla:03}), treating them as AB
magnitudes \citep{Fuk:96}. Of the 791 SF systems, 752 have measurable
[O{\sc ii}] emission.

\begin{figure}
\centerline{\rotatebox{-90}{\includegraphics[width=12cm]{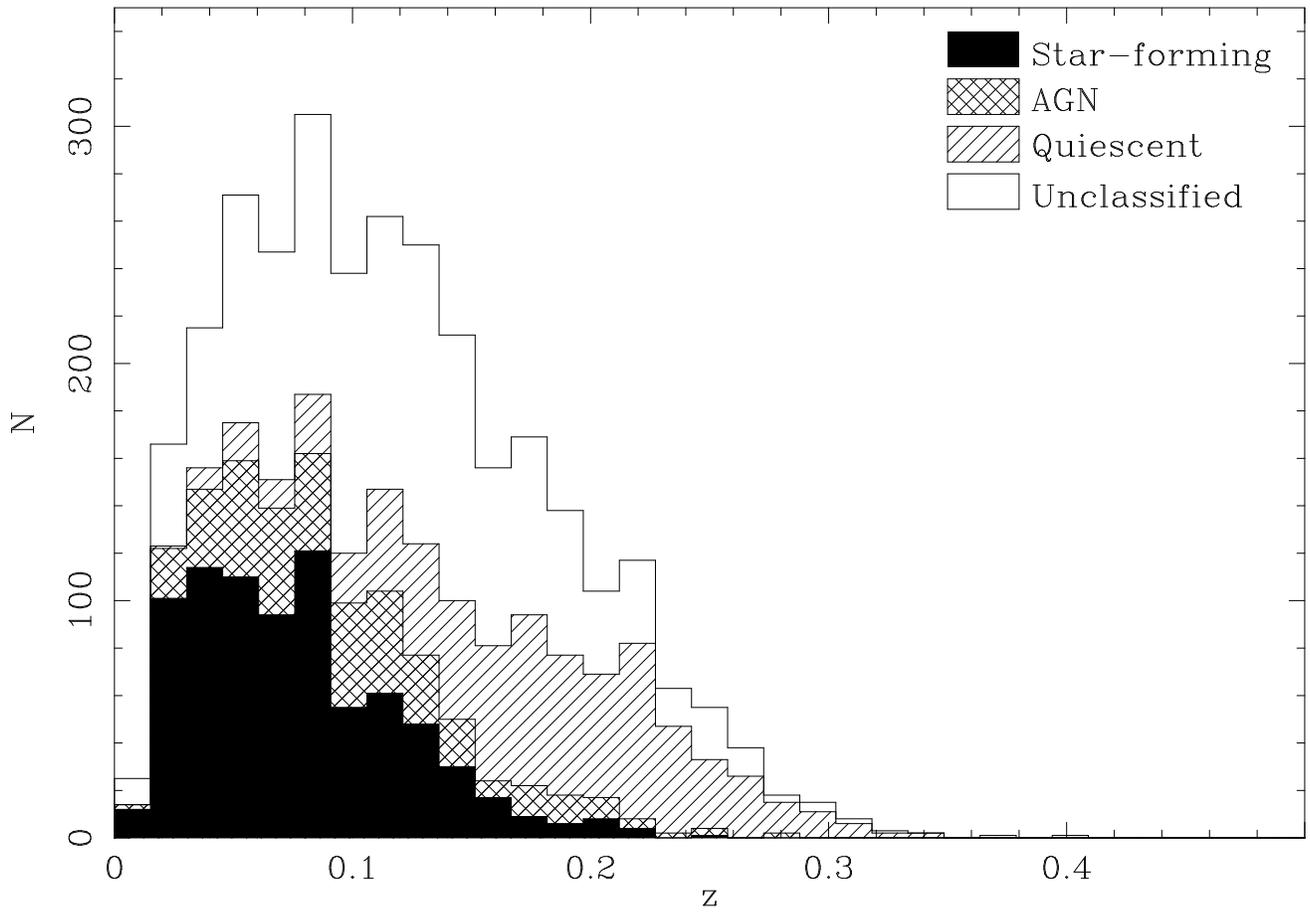}}}
\caption{Redshift distribution of the 1.4\,GHz FIRST detections in
the sample. The histograms indicate the proportions classified as SF, AGN
and ``quiescent." This is done in a cumulative fashion, adding the
histogram for each population onto the previous total, to emphasise the
relative proportions in each bin.
 \label{zhist}}
\end{figure}

\begin{figure}
\centerline{\rotatebox{-90}{\includegraphics[width=12cm]{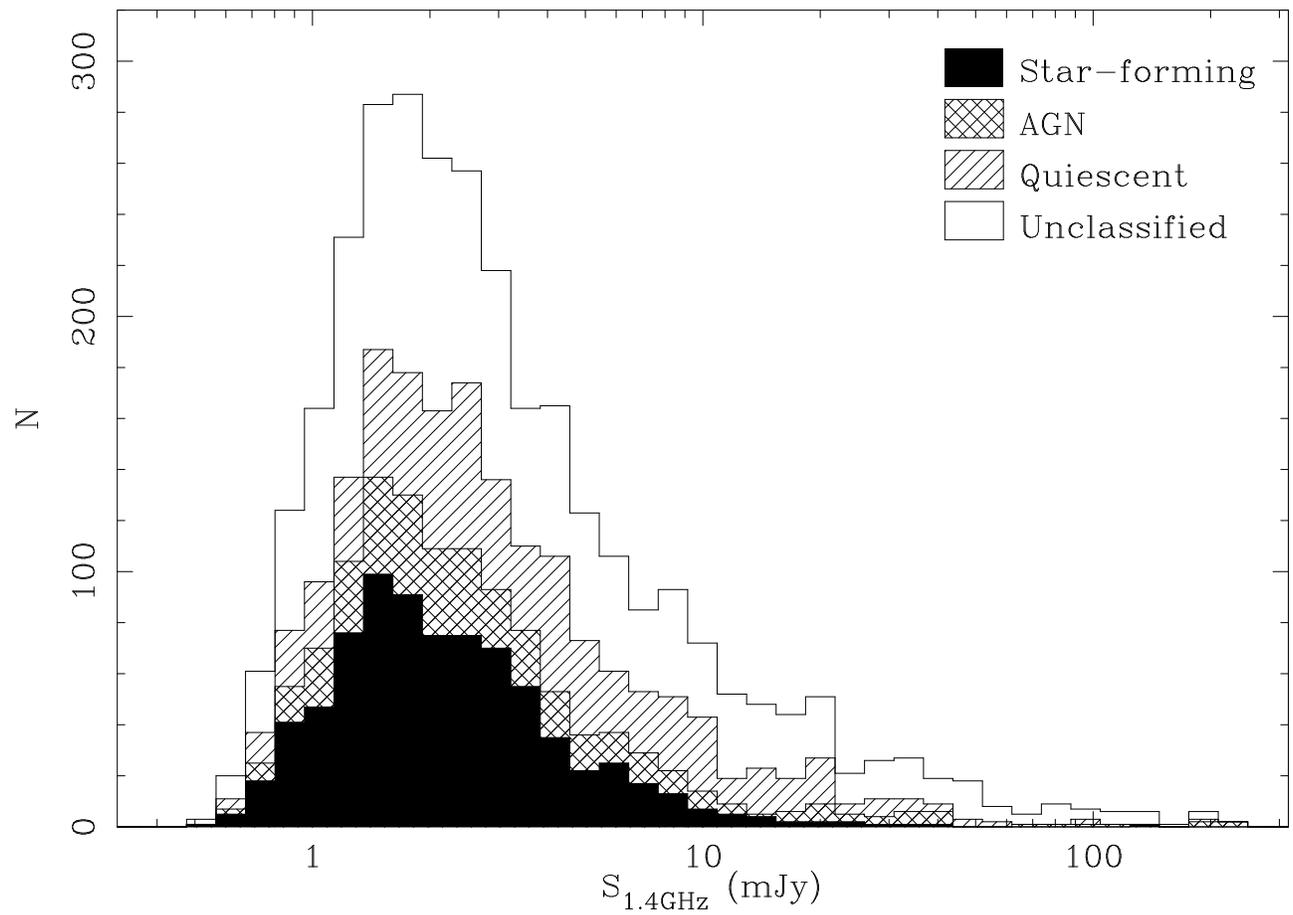}}}
\caption{1.4\,GHz flux density distribution for the FIRST detections in
the sample. As in the previous Figure, the histograms are cumulative,
and indicate the proportions classified as SF, AGN and ``quiescent."
 \label{flxhist}}
\end{figure}

\begin{figure}
\centerline{\rotatebox{-90}{\includegraphics[width=12cm]{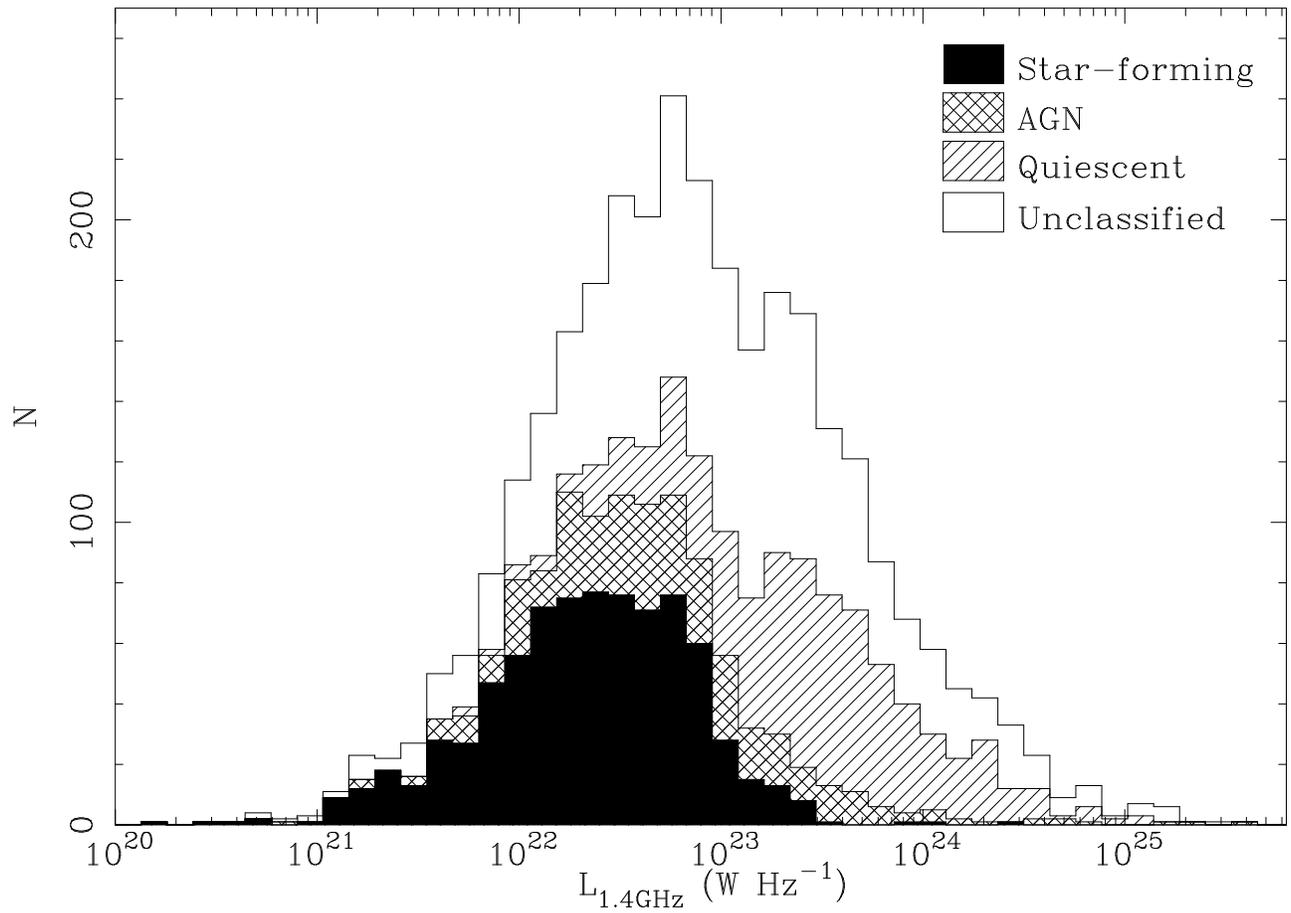}}}
\caption{1.4\,GHz luminosity distribution for the FIRST detections in
the sample. Histograms are again cumulative, indicating the proportions
classified as SF, AGN and ``quiescent."
 \label{lumhist}}
\end{figure}

The completeness of the final spectroscopically classified sample is
complex to define. The initial SDSS spectroscopic sample is complete
to an optical flux limit corresponding to about $r=17.7$, and by
requiring the presence of the [S{\sc ii}] emission lines for classification,
we implicitly impose an upper redshift limit of about $z=0.36$.
By also requiring a 1.4\,GHz detection from FIRST, we impose a second
flux limit, this time corresponding to about 0.5\,mJy. The full
distributions in redshift, 1.4\,GHz flux density and luminosity of
the sample are presented in Figures~\ref{zhist} to \ref{lumhist}.
All three Figures show the distributions for the whole sample of 3079
galaxies, distinguishing the 379 AGNs, 791 SF, and 672 ``quiescent" galaxies.
The proportion of galaxies classified as ``quiescent" becomes more dominant
as redshift increases. This comes about as a combination of the detection
limit of the survey and the necessary diagnostic emission lines moving out of
the observable wavelength window. As a result, some galaxies classified
as ``quiescent," especially those at higher redshifts, may not actually
be quiescent at all. This is emphasised in Figure~\ref{lumhist} where a
large fraction of the brightest radio sources are classified as
``quiescent." In reality these systems may possess strong emission features
that are below the detection threshold or lie outside the accessible SDSS
window. The distribution in 1.4\,GHz luminosities for the 3079 radio
detections spans the very broad range from quite faint SF systems,
$\log(L_{\rm 1.4\,GHz})\approx20.5$, to powerful radio galaxies,
$\log(L_{\rm 1.4\,GHz})\approx26$
\citep[compare with the range found by][for example]{Sad:02}.
The luminosity range spanned by the spectroscopically classified
AGNs is not too dissimilar from that of the SF systems, and this is a
result of two effects: (1) requiring the presence of the necessary
emission lines for the spectroscopic diagnostics limits the classified
AGNs to lower redshifts, and hence lower luminosities; and (2) the more
powerful AGNs may be dominated by elliptical type host galaxies, with
absorption spectra.

Since the results of our SFR analysis do not rely on having a complete
sample, and in order to retain the maximum number of objects for the analysis,
we have chosen not to exclude any galaxies other than those required by
our selection criteria above. Although our full spectroscopically classified
sample thus remains incomplete at the higher redshift and lower flux density
extremes, we can still define regimes over which our sample does approach
completeness. The initial spectroscopic sample is known to be highly
complete over the redshift range $0.05<z<0.1$ \citep{Gom:03}. The flux
density limit to which the FIRST survey is complete is about 2\,mJy
\citep{Whi:97}, so out to $z=0.1$, the 1.4\,GHz detections should be
complete above about $\log(L_{\rm 1.4\,GHz}/{\rm W\,Hz^{-1}})=22.5$
(corresponding, using the calibration described below, to
SFR$_{\rm 1.4\,GHz} \approx 20\,M_{\odot}\,$yr$^{-1}$). Despite this
fairly high lower limit for completeness in SFR, the FIRST detections extend
below the completeness limit to flux densities of about 0.75\,mJy, allowing
SFRs of order unity to be probed, although the radio detected sample at
these levels is highly incomplete.

\section{Star formation rates}
\label{sfr}

Estimating a galaxy's current SFR is typically done by applying a scaling
factor to a star formation sensitive luminosity measurement for the galaxy
\citep{Ken:98,Con:92}. Here we use this technique to estimate SFRs
using H$\alpha$, [O{\sc ii}], $u$-band, 1.4\,GHz and FIR luminosities.
In all the SFR calibrations given below we assume a Salpeter initial mass
function (IMF) and a mass range from 0.1 to 100\,$M_{\odot}$. Changing
the mass range or choosing a different IMF will, of course, alter the
values for the derived SFRs, and a discussion of these effects can
be found in \citet{Ken:98}.
The radio luminosities have been calculated assuming
a power-law spectral index $S_{\nu}\propto\nu^{-0.8}$ appropriate for
galaxies dominated by synchrotron emission, and the optical luminosities are
calculated by incorporating the k-corrections as measured by \citet{Bla:03}.
Below we describe the details of calculating the SFRs for the
five estimators we explore, and comparisons between each are performed.

\subsection{1.4\,GHz Star formation rates}
\label{sfrradio}
Long-wavelength SFR estimates are insensitive to dust obscuration,
increasing their attraction for SFR investigations, but they are not
without limitations. Radio luminosity can be generated by AGN as well as
star formation processes, and indeed the majority of apparently bright radio
sources are AGN. By selecting for star formation directly from optical
spectroscopic features, however, we have eliminated this potential source
of confusion. The detailed physics involved in the connection between SF
and radio emission, in addition, is still poorly understood, despite attempts
by numerous models to explain it \citep[e.g.,][]{Con:92,PD:92,CW:90}.
Despite this, radio luminosities, in part because of the tight correlation
with FIR luminosities \citep[e.g.,][]{deJ:85,Con:91,Yun:01}, appear to be
robust and efficient SFR estimators \citep{Con:92,Cram:98,Bell:03}. This,
combined with the fact that the 1.4\,GHz luminosity is insensitive to dust
obscuration makes the use of SFR$_{\rm 1.4\,GHz}$ very attractive, and we
adopt this as a reference SFR when investigating the details of the other
SFR measures. After the detailed investigation of the radio-FIR correlation
by \citet{Bell:03}, we adopt the calibration derived therein between
1.4\,GHz luminosity and SFR,
\begin{equation}
\label{sfr1.4}
{\rm SFR_{1.4GHz}}\,(M_{\odot}\,{\rm yr^{-1}}) =
 \frac{f L_{\rm 1.4GHz}}{1.81 \times 10^{21}\,{\rm W\,Hz^{-1}}},
\end{equation}
where
\begin{equation}
f = 
\left\{
\begin{array}{ll}
 1 & L_{\rm 1.4GHz} > L_c \\
 (0.1 + 0.9 (L_{\rm 1.4GHz}/L_c)^{0.3})^{-1} & L_{\rm 1.4GHz} \le L_c,
\end{array}
\right.
\end{equation}
and $L_c=6.4\times10^{21}\,$W\,Hz$^{-1}$. This calibration produces SFRs
about a factor of two lower than the calibration of \citet{Con:92}
for luminosities above $L_c$, while for fainter luminosities the
SFRs progressively converge, with the present calibration eventually
producing larger SFRs than that of \citet{Con:92}
below about $3.4\times10^{20}\,$W\,Hz$^{-1}$
\citep[see discussion in][]{Bell:03}.

\begin{figure}
\centerline{\rotatebox{-90}{\includegraphics[width=12cm]{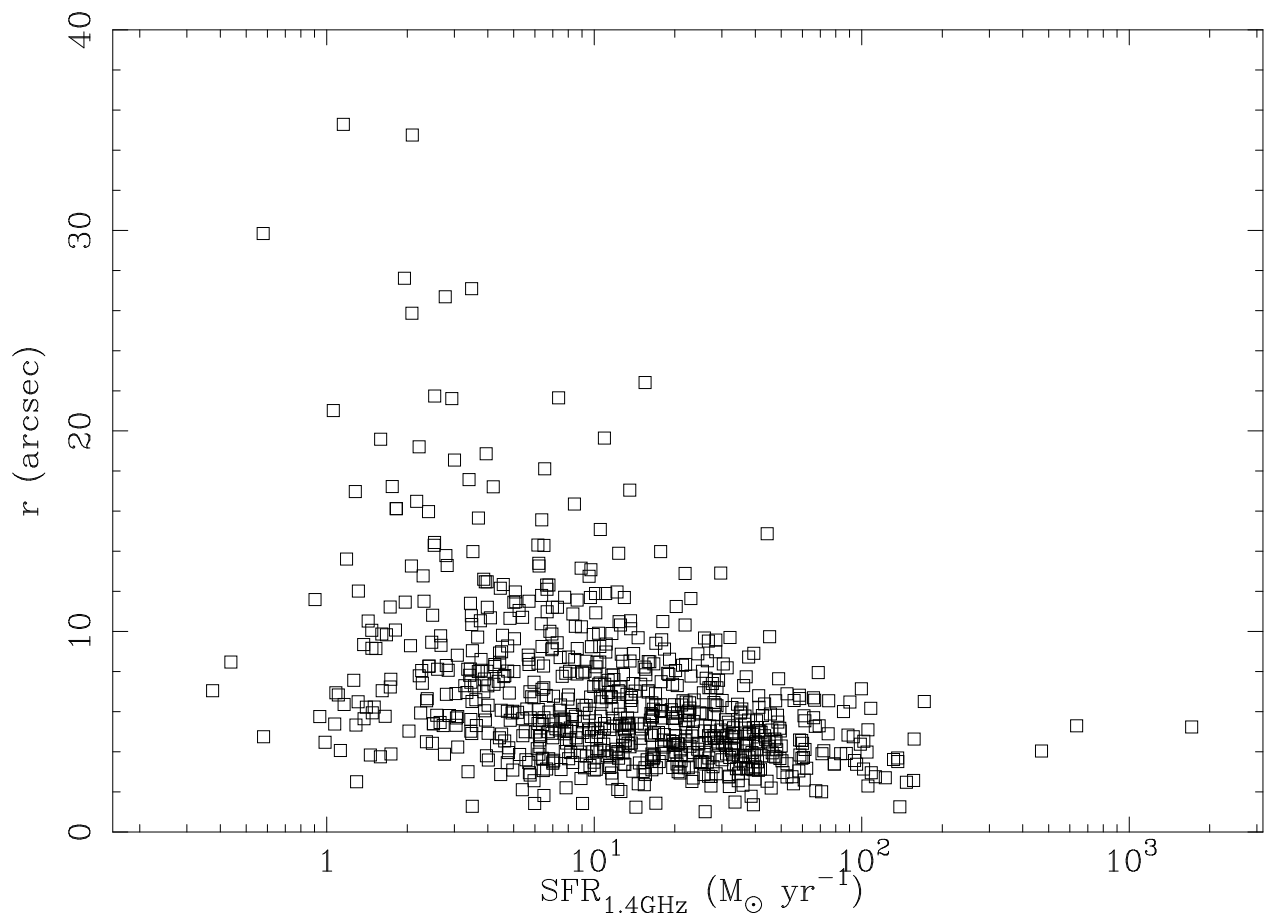}}}
\caption{Galaxy size given by the Petrosian radius in $r$-band
as a function of SFR$_{\rm 1.4GHz}$. Here it can be explicitly seen that
the FIRST SFRs below about $10\,M_{\odot}\,$yr$^{-1}$ belong to
progressively larger galaxies, implying that these SFRs are progressively
underestimated.
 \label{sizvssfr}}
\end{figure}

\begin{figure}
\centerline{\rotatebox{-90}{\includegraphics[height=12cm]{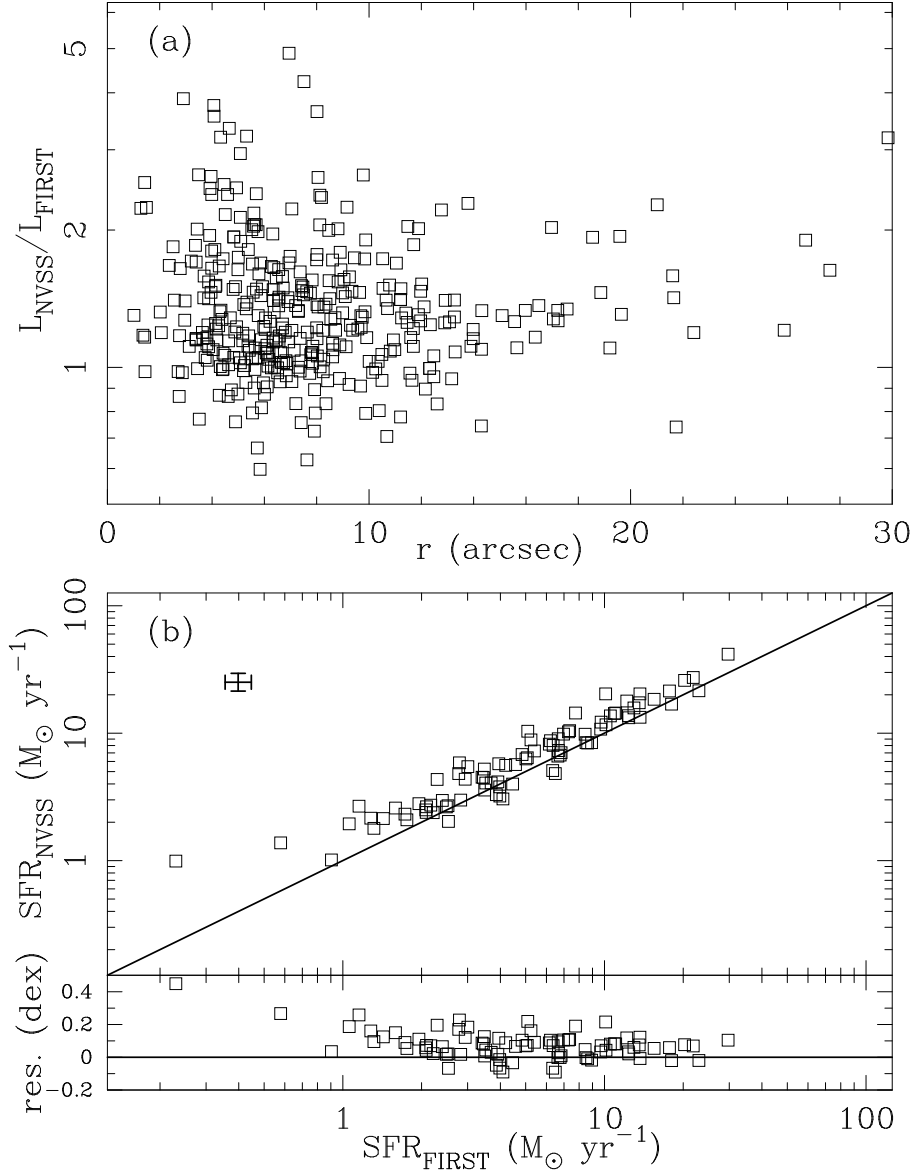}}}
\caption{(a) Ratio of NVSS to FIRST luminosities for SF galaxies as a
function of Petrosian radius in $r$-band. This shows that the majority of
sources smaller than about $r=10''$ have comparable luminosities (and flux
densities) from NVSS and FIRST, although NVSS still appears to be fractionally
larger (by about 5\% to 10\%) on average. Above $10''$, though, the NVSS flux
density is consistently higher than that from FIRST. There are a handful of
sources with $r<10''$ with significantly higher NVSS flux densities, and these
are likely to be cases where one or more nearby galaxies have entered the NVSS
beam in addition to the galaxy for which the FIRST flux density is measured.
(b) Comparison of the SFRs derived using NVSS and FIRST 1.4\,GHz
luminosities, for all SF galaxies having $r>10''$. The
greater sensitivity of the NVSS to extended radio emission is clearly
seen in the comparison. The error bars in the upper left indicate the typical
uncertainty in the measurements. They represent random error only, no
systematic errors (such as the uncertainty in the SFR calibration) are
included. The lower panel shows the residuals from the one-to-one line.
 \label{radsfrcomp}}
\end{figure}

Most of the current generation of sensitive radio surveys, including FIRST,
are conducted with interferometric telescopes, which, lacking short-spacing
information, have a limited sensitivity to emission from more extended
structures. The extent of this effect is dependent on the shortest baseline of
the instrument. In particular, the FIRST catalogue starts losing sensitivity
to 1.4\,GHz emission for galaxies larger than about $10''$. For galaxies with
a radius of $12''$ FIRST is only sensitive to about 84\% of the emission,
decreasing further for larger sources \citep{Bec:95}. If this effect is
not accounted for, it can result in an apparent overestimate of the
comparison SFR at low SFRs \citep[see, for example, Figure~1 of][]{Cram:98},
although it is actually an underestimate in the 1.4\,GHz derived SFRs.
As shown in Figure~\ref{sizvssfr}, where the galaxy size is shown as
a function of SFR$_{\rm 1.4\,GHz}$ from FIRST, this starts becoming an
issue in the current sample for galaxies with
SFR$_{\rm 1.4\,GHz} \lesssim 10\,M_{\odot}\,$yr$^{-1}$. To avoid such
underestimates in SFR$_{\rm 1.4\,GHz}$ for our analysis,
we make use of the NRAO VLA Sky Survey \citep[NVSS,][]{Con:98},
a 1.4\,GHz survey made with the VLA in the more compact D configuration.
As a result of the compact VLA configuration used for the survey, the
NVSS has poorer resolution than FIRST, but greater sensitivity
to extended structure. It also has a survey limit similar to FIRST, and
of the 107 SF galaxies in our sample with $r>10''$, 87 are present
in the NVSS catalogue, within a $15''$ matching radius of the SDSS object
\citep[c.f.\ the $10''$ radius used by][for NVSS and 2dFGRS matches]{Sad:02}.
These objects are shown in Figure~\ref{radsfrcomp}, where the NVSS 1.4\,GHz
luminosities are compared with those from FIRST. For the systems larger than
about $10''$ the extent of the underestimate in the FIRST measurements is
clear. For all subsequent analysis herein, we use the NVSS
derived SFRs in place of those from FIRST for galaxies with $r>10''$ to
ensure no underestimates of 1.4\,GHz derived SFR bias our results. Omitting
the 20 objects with $r>10''$ and no NVSS measurement leaves a final sample
of 771 SF galaxies.

\subsection{H$\alpha$ Star formation rates}
\label{hasfr}
In estimating SFR from the H$\alpha$ luminosity we adopt the calibration
given by \citet{Ken:98}:
\begin{equation}
\label{sfrha}
{\rm SFR_{H\alpha}}\,(M_{\odot}\,{\rm yr^{-1}}) =
   \frac{L_{\rm H\alpha}}{1.27\times10^{34}\,{\rm W}}.
\end{equation}
Prior to applying this calibration, though, there are several issues
to address regarding the measurement of an H$\alpha$ luminosity representative
of the full emission from a galaxy given a flux measurement from a
fiber-based spectrum. These issues include both corrections for obscuration,
due to the intrinsic dust content of the target galaxy, and aperture
corrections, to account for the emission missed by virtue of the fiber
diameter potentially being smaller than a target galaxy.
Both these effects can be accounted for using the available SDSS data.

Before addressing these issues, we briefly digress to comment on the
observing strategy called ``smearing" \citep{Sto:02}. This is used in the
SDSS to account for seeing and wavelength-dependent atmospheric refraction
effects in ensuring the most accurate spectrophotometric calibration.
A ``smear-correction," in brief, consists of scaling a primary spectrum so
its smoothed continuum equals the smoothed continuum of a short exposure of
the same object taken while the telescope pointing was dithered to cover a
somewhat larger aperture than the fiber diameter. This scaling preserves line
equivalent widths, but not line fluxes. The line fluxes are effectively being
``aperture corrected" by scaling them assuming the line emission scales
directly with the stellar continuum. This actually follows the same
principle as the methods used below to perform aperture corrections,
although the scaling of the smear-correction is to the stellar continuum
within a fixed aperture for all galaxies, and will not necessarily
encompass all of a target galaxy, especially at lower redshifts.

Naively using measurements of spectral line fluxes from smear-corrected
spectra can thus potentially lead to small errors in the resulting fluxes
and line ratios. Because of the fixed aperture effectively used in
constructing the smear-correction, the line fluxes from smear-corrected
spectra will be different from those constructed from the primary spectra
with an aperture correction based on measured galaxy sizes or magnitudes.
Some preliminary comparisons for line ratios, using Balmer decrement
measurements in SF galaxies, quantify this small effect to some extent.
Even with the Balmer decrement line ratio (widely spaced in wavelength)
the systematic difference between primary and smear-corrected spectra
is only about 10\% at most.

In summary, equivalent widths measured from smear-corrected spectra should
be consistent with those in primary spectra, and line ratios for lines close
to each other in wavelength (such as the typical AGN/SF diagnostics)
will be negligibly affected. But it is clear that to avoid potentially
introducing small systematic offsets, the best solution for individual flux
measurements, or line ratios widely separated in wavelength, is to use spectral
line measurements made on spectra which have been processed without using the
smear correction. All spectral line measurements used for the analysis herein
were made on primary spectra, with no smear corrections applied.

\subsubsection{Obscuration corrections}
\label{obs}

Galactic foreground obscuration is corrected for following \citet{Sch:98}, but
obscuration by dust intrinsic to the SF galaxies can cause more significant
underestimates in the emission line and $u$-band derived SFRs. We address this
by making obscuration corrections in two ways. All the SF classified galaxies
have measured H$\alpha$ and H$\beta$ fluxes, so the Balmer decrement
F$_{\rm H\alpha}$/F$_{\rm H\beta}$ can be calculated, and used to estimate
and correct for the obscuration. The suggestion of luminosity-dependent
obscuration in SF galaxies \citep[e.g.,][]{Hop:01,Sul:01}, also recently
identified in individual regions of star formation from spatially
resolved spectroscopy of an extreme SF galaxy, IRAS~19254-7245
\citep{Ber:03}, is explored as well. New results from the
Phoenix Deep Survey \citep{Afo:03} indicate, however, that the
situation is more complex than implied by simple linear
models. While galaxies with low SFRs seem to have relatively low
levels of obscuration, at higher SFRs a broad range of obscurations are
seen. A trend for an increase in the median obscuration with SFR, though,
is still present. We make use of the method described by \citet{Afo:03} and
the relationship they derive to explore the effectiveness of such an
empirical correction for the present sample, and perform an explicit
comparison with the Balmer decrement correction.

Stellar absorption in the Balmer emission lines, if not accounted for,
can cause a significant overestimation in the implied obscuration from
measurements of the Balmer decrement \citep{Ros:02}. Detailed analysis
of continuum fitting to refine emission line measurements is being
addressed by \citet{Tre:03}, and an alternative line measurement method
explored by \citet{Goto:03}, but for the purposes of the current
investigation it was deemed sufficient to assume a simple constant
correction for stellar absorption in the measured Balmer line
equivalent widths (EWs). The value of this equivalent width correction,
${\rm EW_c}$, for the H$\beta$ line was found by \citet{MO:02} to vary 
from about 1\,\AA\ for Sa galaxies to about 4\,\AA\ for extreme late types,
with ${\rm EW_c}=2\,$\AA\ for Sb galaxies, consistent with the typical value
found in other studies of SF galaxies \citep[e.g.,][]{Age:99,Tres:96}.

Given the broad range of possible stellar absorption
values, the assumption of a common value may act to introduce a degree of
scatter into the resulting SFR estimates. For the sample investigated here, the
stellar absorption correction is typically a significant fraction of the
measured H$\beta$ equivalent width, anything from a 10\% to 100\% correction.
As a result, it is worth emphasising that the extent of the derived obscuration
correction depends strongly on the assumed value of the stellar absorption.
The assumption of a common value as done here
should therefore be restricted only to studies of large samples where
the gross characteristics of the population are being examined, and more
refined measurements should be preferred for analyses of individual objects.
Furthermore, since the relative correction is largest for systems with
low EW, and since these tend to be those with higher luminosities, it is
the bright galaxies that are most affected by any uncertainty in the
extent of the absorption correction.

The EW correction is converted into a flux correction for the Balmer
lines using
\begin{equation}
\label{stelabs}
S = \frac{\rm EW + EW_c}{\rm EW} F,
\end{equation}
where $S$ is the stellar absorption corrected line flux for H$\alpha$ or
H$\beta$, EW is the equivalent width of the line, ${\rm EW_c}$ the correction
for stellar absorption, and $F$ the observed line flux. The commonly assumed
stellar absorption of ${\rm EW_c}=2\,$\AA\ at H$\beta$ derives from a line
measurement technique which involves an integration of the spectrum over a
wavelength range encompassing the line \citep{MO:02,Goto:03}. When
alternative measurements are used, such as the Gaussian fitting of the SDSS
pipeline, it appears that a smaller stellar absorption correction is necessary.
This is a result of the SDSS spectral resolution, which partially resolves the
stellar absorption, resulting in the fitted Gaussian being only incompletely
diminished by the absorption. Thus the measured line flux (or EW) is larger
than a flux-summing method would produce.

\begin{figure}
\centerline{\rotatebox{0}{\includegraphics[width=17cm]{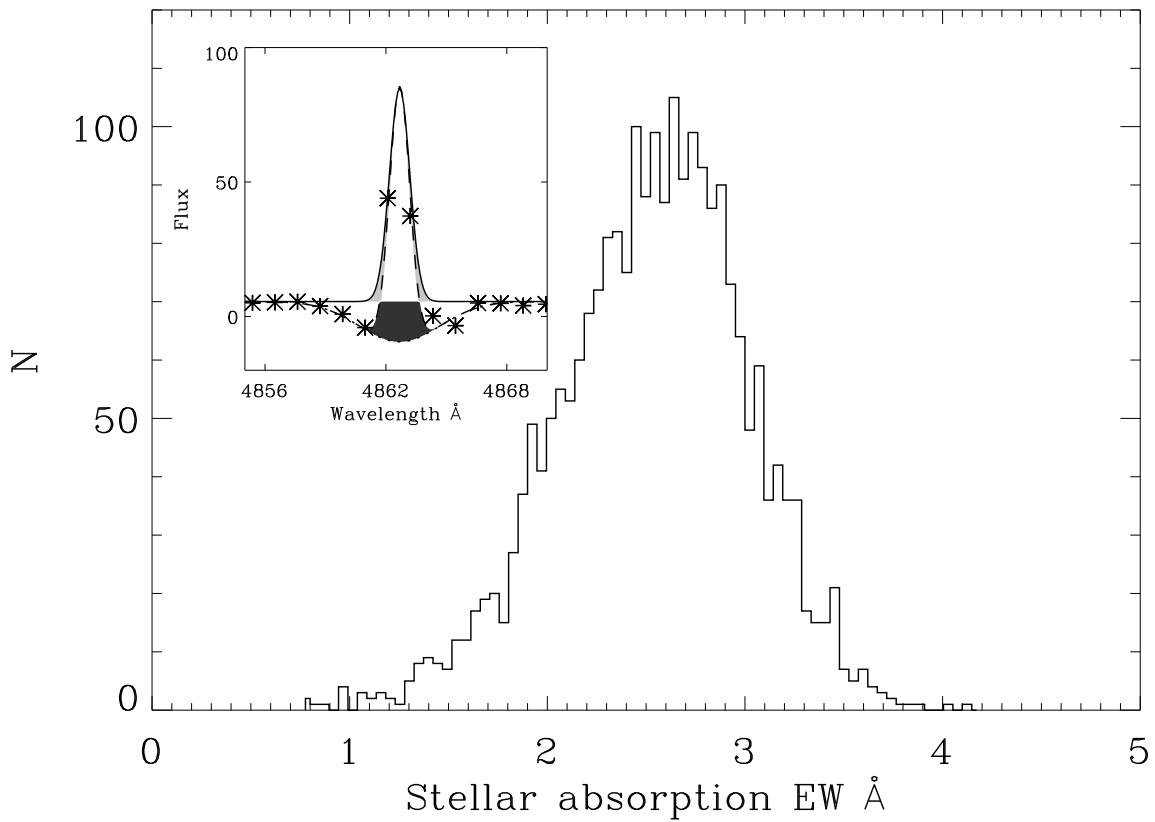}}}
\caption{Histogram of intrinsic stellar absorption EWs at H$\alpha$
for the complete sample of SF galaxies. The median value is 2.6\,\AA.
Inset: Sample SDSS spectrum indicating the effect of the pipeline
measurement of the H$\beta$ emission line, illustrating the incomplete
extent to which stellar absorption reduces the measured emission. The
points are the actual measured SDSS spectrum.
The solid line is the SDSS pipeline Gaussian fit to the emission, and
the dashed lines are a double Gaussian fit, made by fixing an absorption
component to have an EW$=3.0\,$\AA\ the same EW as the absorption measured
at H$\delta$ in this spectrum. The black shaded region indicates the
true flux missed by the pipeline Gaussian fit in the measurement of
the emission, while the grey shaded regions indicate excess flux added by
the pipeline Gaussian fit.
The black area minus the grey area is the true amount by which the
pipeline fit underestimates the emission. It can clearly be seen that this
area is smaller (by at least a factor of two) than the area of the
stellar absorption component.
Given a median stellar absorption of EW$=2.6\,$\AA\ at H$\alpha$,
the SDSS pipeline measurements, typically, are diminished by only
about EW$=1.3\,$\AA.
 \label{stelabscor}}
\end{figure}

To establish the extent of this effect, the approximate distribution of actual
stellar absorptions was first explored to determine the median absorption at
the wavelength of H$\alpha$. Rather than simply assuming ${\rm EW_c}=2\,$\AA\
for H$\beta$, we use the measured absorption at H$\delta$ in the SDSS spectra
as a proxy for the H$\beta$ absorption. We then followed the prescription of
\citet{Keel:83} for deriving the stellar absorption at H$\alpha$,
EW(H$\alpha$)$=1.30+0.40\,$EW(H$\beta$) \citep{MO:02}. The resulting
distribution of stellar absorption EWs at H$\alpha$ for a complete, volume
limited sample of SF galaxies (constructed below, see \S\,\ref{sectionsfru})
is shown in Figure~\ref{stelabscor}. The median value of this distribution is
2.6\,\AA, consistent with the observed range for late type galaxies. Since the
resolution of SDSS spectra are sufficient to resolve this absorption, a
correction smaller than this actual value needs to be applied. This is
demonstrated explicitly for an actual SDSS spectrum in the inset of
Figure~\ref{stelabscor}. The SDSS pipeline Gaussian fit to the emission line
underestimates the line flux by the black shaded area minus the grey shaded
area, which can be seen to be smaller than the total stellar absorption, by
at least a factor of two. Given the median stellar absorption of $2.6\,$\AA,
it seems that the appropriate stellar absorption correction should be at most
about ${\rm EW_c}=1.3\,$\AA. We adopt this value for the rest of this analysis.

Now, using the stellar absorption corrected Balmer line measurements, the
Balmer decrement $S_{\rm H\alpha}/S_{\rm H\beta}$ can be constructed.
The obscuration correction is then derived using the Balmer decrement
and an obscuration curve. For making obscuration corrections to the
emission lines, we use the Milky Way obscuration curve of \citet{Car:89}
as referenced in Table~2 of \citet{Cal:01}. For the obscuration corrections
to the stellar continuum in the $u$-band (\S\,\ref{sectionsfru}), we use the
obscuration curve of \citet{Cal:00} derived for starburst galaxies.
Note that the extent of the obscuration experienced by emission lines and
the stellar continuum at the same wavelength differs by a factor of about
two \citep{Cal:01}.

\begin{figure}
\centerline{\rotatebox{-90}{\includegraphics[width=12cm]{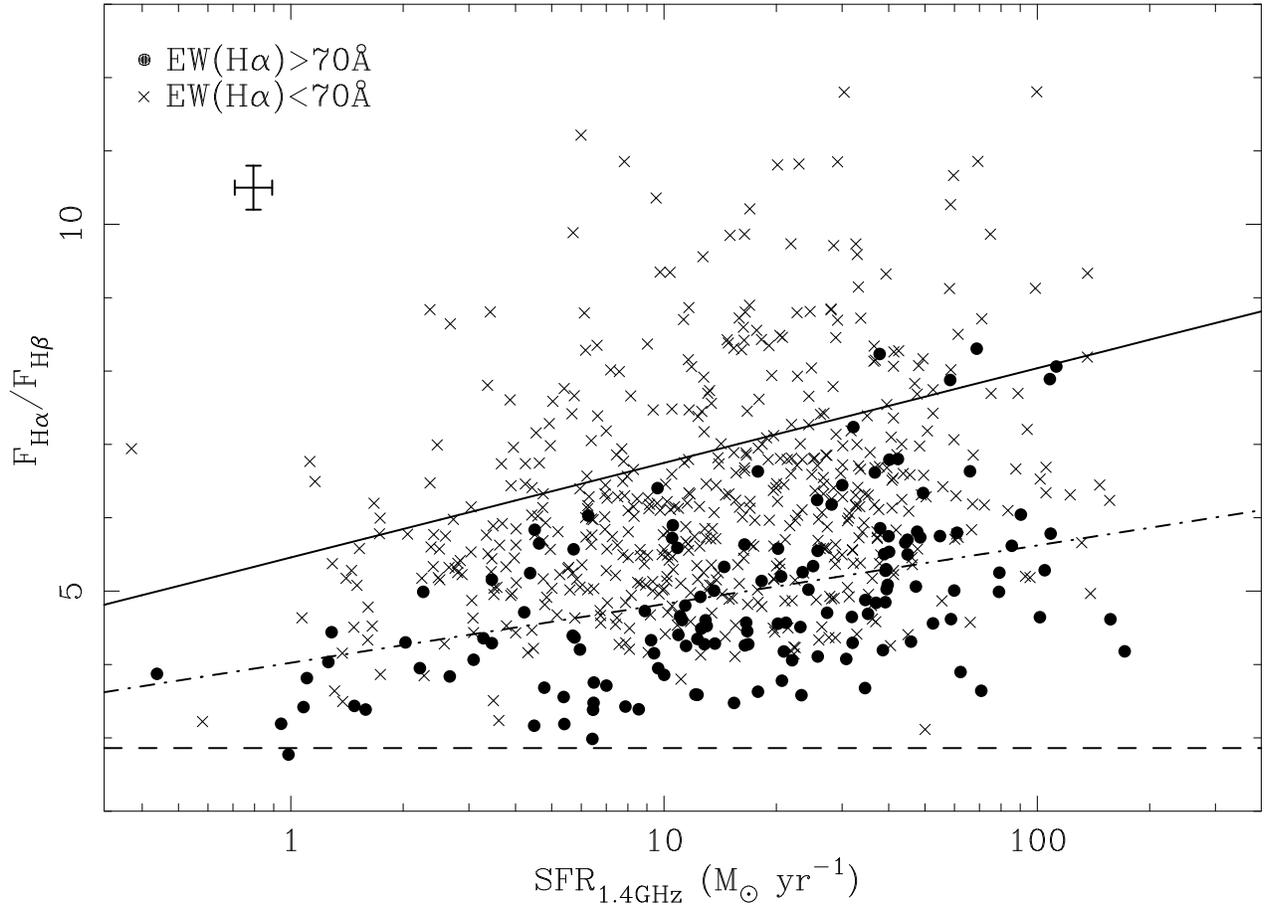}}}
\caption{Balmer decrements, corrected for stellar absorption as
described in the text, as a function of SFR$_{\rm 1.4\,GHz}$.
The different symbols indicate ranges in H$\alpha$ equivalent width,
as shown. The error bars in the upper left indicate the typical uncertainty
in the measurements. They represent random error only, no systematic 
errors (such as the uncertainty in the SFR calibration) are included.
The dashed line indicates the 2.86 value expected from case B
recombination \citep{Bro:71}, the solid line indicates the Balmer
decrements predicted from the SFR-dependent obscuration derived
by \citet{Afo:03}, and the dot-dashed line those from \citet{Hop:01}.
(These relations have been converted to be consistent with the 1.4\,GHz
SFR calibration and cosmology used here.)
While the empirical correction of \citet{Hop:01} is clearly an underestimate
for the sample on average (although it may be reasonable for systems
with EW(H$\alpha)\gtrsim70\,$\AA), that of \citet{Afo:03} seems to be somewhat
of an overestimate, on average, for this sample. This is likely to reflect the
radio-selected nature of the sample from which this relation was derived,
suggesting that the present sample may be missing a number of more highly
obscured systems.
 \label{bdecvssfr}}
\end{figure}

The stellar absorption corrected Balmer decrements are shown as a function of
SFR$_{\rm 1.4GHz}$ in Figure~\ref{bdecvssfr}. The predicted Balmer decrement
from the SFR-dependent obscuration of \citet{Afo:03} is shown as the
solid line in this Figure, and that of \citet{Hop:01} as the dashed line.
The relation independently derived by \citet{Sul:01} lies $10\%$ higher
than that of \citet{Hop:01}, while having an almost identical slope.
These relations have been converted to the current SFR calibrations and
cosmologies where necessary. As discussed by \citet{Afo:03}, the empirical
relation found by \citet{Hop:01} appears to be affected by sample selection
effects (being restricted to higher EW systems). This results in a model
that provides a reasonable approximation for larger EW systems
(EW(H$\alpha$)$\gtrsim70$\,\AA), but is clearly an underestimate for
smaller EW systems. For the whole sample, the model of \citet{Afo:03}
is better at tracing the trend in the median obscuration with SFR, although 
appears to be a slight overestimate. This is discussed further in
\S\,\ref{hacomp} below. Again, very importantly, there is significant scatter
in the distribution of actual Balmer decrements about such trends,
particularly at high SFR.

On top of the overall trend for systems with higher SFRs to have a higher
median Balmer decrement, the observed trend with EW in Figure~\ref{bdecvssfr}
is also likely to be real, and not merely an artifact of the chosen stellar
absorption correction. In the relatively low SFR range
$1<{\rm SFR_{1.4GHz}}<10\,M_{\odot}\,$yr$^{-1}$, for example,
the systems with EW$<70\,$\AA\ have a median Balmer decrement of 6.1,
compared with 5.3 for the systems with larger EWs. If the stellar absorption
correction was boosted as high as  ${\rm EW_c}=2.6\,$\AA, (an
unreasonably large estimate given the illustration in Figure~\ref{stelabscor}),
these median values decrease only to 5.4 and 4.6 respectively. This indicates
both that there are real differences in the extent of the typical absorption,
depending on the observed EW, and that this difference is not an artifact
of the chosen stellar absorption correction. The extent of the absorption,
moreover, is not negligible, even at these relatively low SFRs.

\subsubsection{Aperture corrections}
\label{apcor}

In addition to the obscuration correction, the emission line luminosities
also require an aperture correction to account for the fact that only
a limited amount of emission from a galaxy is detected through the
$3''$ diameter fiber. For both the H$\alpha$ and [O{\sc ii}] emission lines,
this is done as follows.

The H$\alpha$ EW (corrected for stellar absorption) can be used along
with an estimate of the continuum luminosity for the galaxy from the
photometric catalogue at the observed wavelength of H$\alpha$, to recover
an effective H$\alpha$ line luminosity for the whole galaxy. Explicitly,
(before obscuration corrections are applied),
\begin{equation}
\label{ewsfr}
L_{\rm H\alpha}\,({\rm W}) = ({\rm EW+EW_c})\,10^{-0.4(M_{\rm r}-34.10)}\,
 \frac{3\times10^{18}}{(6564.61(1+z))^2}
\end{equation}
where $M_{\rm r}$ is the k-corrected absolute $r$-band AB-magnitude, derived
from the observed $r$-band Petrosian magnitude. The last term converts this
luminosity from units of W\,Hz$^{-1}$ to W\,\AA$^{-1}$.
This Equation assumes that the flux of the continuum at the wavelength of
H$\alpha$ can be represented by the flux at the effective wavelength of
the $r$-band filter ($\approx 6222\,$\AA). While this is not strictly true,
the continuum in these SF systems in the wavelength range of interest is
flat enough that it is a good approximation. A more refined estimate can be
made using an interpolation between the absolute magnitudes in the
$r$ and $i$, or $i$ and $z$ filters, as appropriate for the redshift of the
galaxy. This was explored and found to make a negligible difference
in the resulting distribution of SFRs and in the comparison with SFRs
estimated at other wavelengths.

The aperture correction implicitly assumes that the
emission measured through the fiber is characteristic of the whole galaxy,
and that the SF is uniformly distributed over the galaxy. Issues such as
galaxy orientation and patchy distributions of SF regions, primarily when the
aperture corrections are large, will increase the uncertainties in this
form of aperture correction estimate.

In the case of the [O{\sc ii}] luminosity, an exactly analogous
method is used to calculate the aperture correction, although no
stellar absorption correction is necessary for the EW of [O{\sc ii}].
In Equation~\ref{ewsfr} $M_{\rm u}$ is substituted for $M_{\rm r}$,
and the wavelength of [O{\sc ii}], 3728.30\,\AA, is used in place of
that of H$\alpha$, 6564.61\,\AA, in the last term. The explicit calculation
used in constructing the final SFR$_{\rm H\alpha}$, incorporating both the
aperture correction and the obscuration correction, is given as
Equation~\ref{apobssfr} in Appendix~\ref{formulae}.

An alternative method of applying an aperture correction is described
in Appendix~\ref{apcor2}, along with a fuller discussion of systematic
uncertainties and an exploration of how the aperture correction depends
on parameters such as angular size and redshift.

\subsubsection{Comparison with 1.4\,GHz SFRs}
\label{hacomp}
\begin{figure}
\centerline{\rotatebox{-90}{\includegraphics[width=12cm]{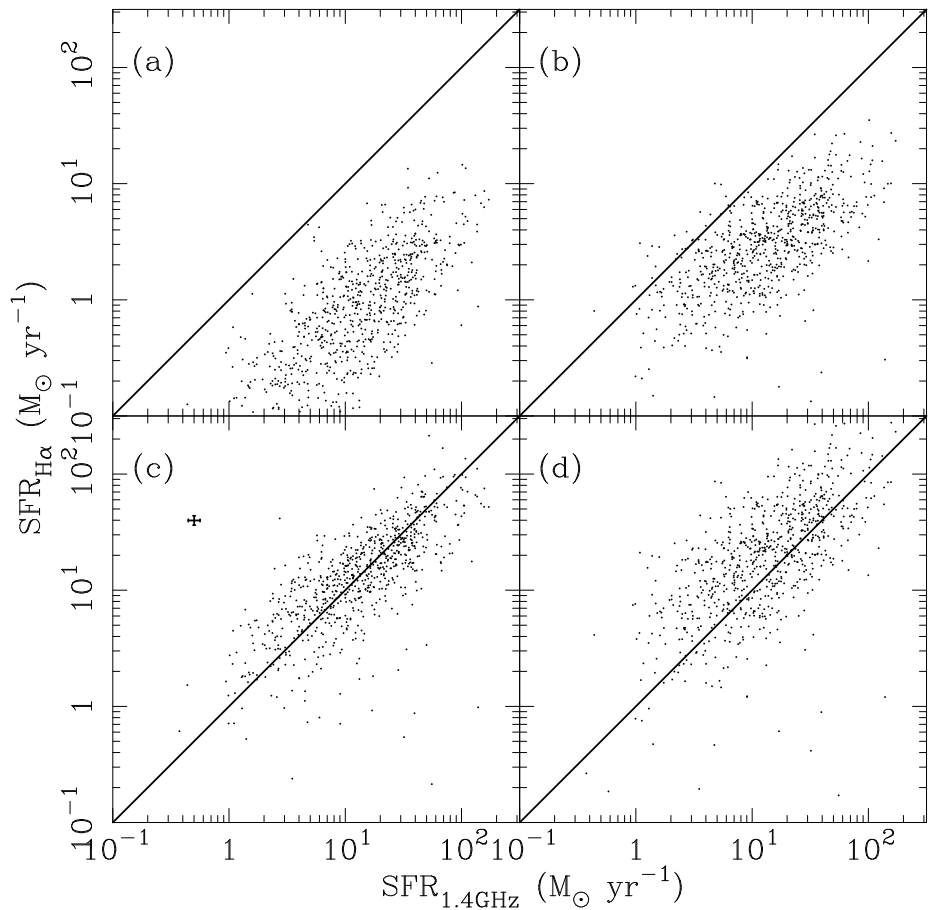}}}
\caption{SFRs from H$\alpha$ luminosity as a function of SFRs from FIRST
1.4\,GHz luminosity. Only galaxies spectroscopically classified as SF are
shown. All panels show the 771 SF galaxies
remaining when NVSS measurements are used for galaxies having $r>10''$.
Panel (a) shows SFR$_{\rm H\alpha}$ uncorrected for aperture or obscuration
effects. Panel (b) shows the aperture corrected SFR$_{\rm H\alpha}$ prior to
the obscuration correction, calculated from Equation~\ref{ewsfr}, and
panel (c) adds the obscuration correction using the Balmer decrement to
give the fully corrected SFR$_{\rm H\alpha}$. Panel (d) uses the method
of \citet{Afo:03} rather than the Balmer decrement for making the
obscuration correction, for comparison. The rms deviation either side
of the one-to-one line in (c) is 0.21 dex, a factor of 1.6.
The error bars in the upper left of panel (c) indicate the typical
uncertainty in the measurements. Again, they include random error only.
 \label{radhasfr}}
\end{figure}

The estimated SFRs derived from H$\alpha$ luminosities are shown as
a function of SFR$_{\rm 1.4\,GHz}$ in Figure~\ref{radhasfr}.
Each step in the process of calculating the H$\alpha$ luminosities is
shown, to emphasise the magnitude of each effect. The uncorrected SFRs
shown are calculated directly from the H$\alpha$ line luminosities
prior to any obscuration or aperture corrections. The aperture corrected
SFRs are calculated from the luminosities obtained using Equation~\ref{ewsfr}
before applying obscuration corrections, and finally, from
Equation~\ref{apobssfr} incorporating the obscuration correction as well.
It can be seen that, on average, the combined effect of the aperture and
obscuration correction is to increase uncorrected luminosities (and SFRs)
by a factor of about 20. The resulting estimates from H$\alpha$ and 1.4\,GHz
luminosities are highly consistent, although there remains a significant
amount of scatter. The rms characterising the extent of this scatter,
in the sense of the rms deviation from the one-to-one line, is 0.21 dex,
or about a factor of 1.6 either side of the line. The extrema of the
scatter (apart from a small number of significant outliers) span about 1 dex.

The results of using the obscuration correction method of \citet{Afo:03}
are also shown in Figure~\ref{radhasfr}, producing a similar distribution for
the derived H$\alpha$ SFRs as the Balmer decrement correction.
Directly comparing SFR$_{\rm H\alpha}$ estimated using obscuration corrections
derived from the Balmer decrement with those using the method of
\citet{Afo:03} show that the relation of \citet{Afo:03} gives, on average,
an overestimate of about 20\%. This is not unexpected given the location of
the trend shown in Figure~\ref{bdecvssfr}. \citet{Afo:03} discusses the
fact that this relation, derived from a radio-selected sample of SF galaxies,
reflects the presence of galaxies with larger obscurations, able to be
detected in radio-selected samples, and possibly overlooked in UV or
optically selected samples. This point will be discussed in more detail in
\S\,\ref{disc_rad} below. In any case, for this type of application
empirical relations like that of \citet{Afo:03} and \citet{Hop:01},
when their selection effects and limitations are correctly taken into
account, appear to be useful tools for estimating obscuration corrections
in the absence of more physical measurements.

It is possible that some of the scatter observed in Figure~\ref{radhasfr}
may be produced by our assumption of a common stellar absorption correction.
This was investigated by applying corrections using the individual estimates
derived above, from the method of \citet{Keel:83}. No measurable reduction in
the resulting scatter was detected, suggesting that the assumption of
a common EW correction for stellar absorption does not dominate the observed
scatter in these results. To account for the scatter, other physical processes
must be investigated.

\begin{figure}
\centerline{\rotatebox{-90}{\includegraphics[width=12cm]{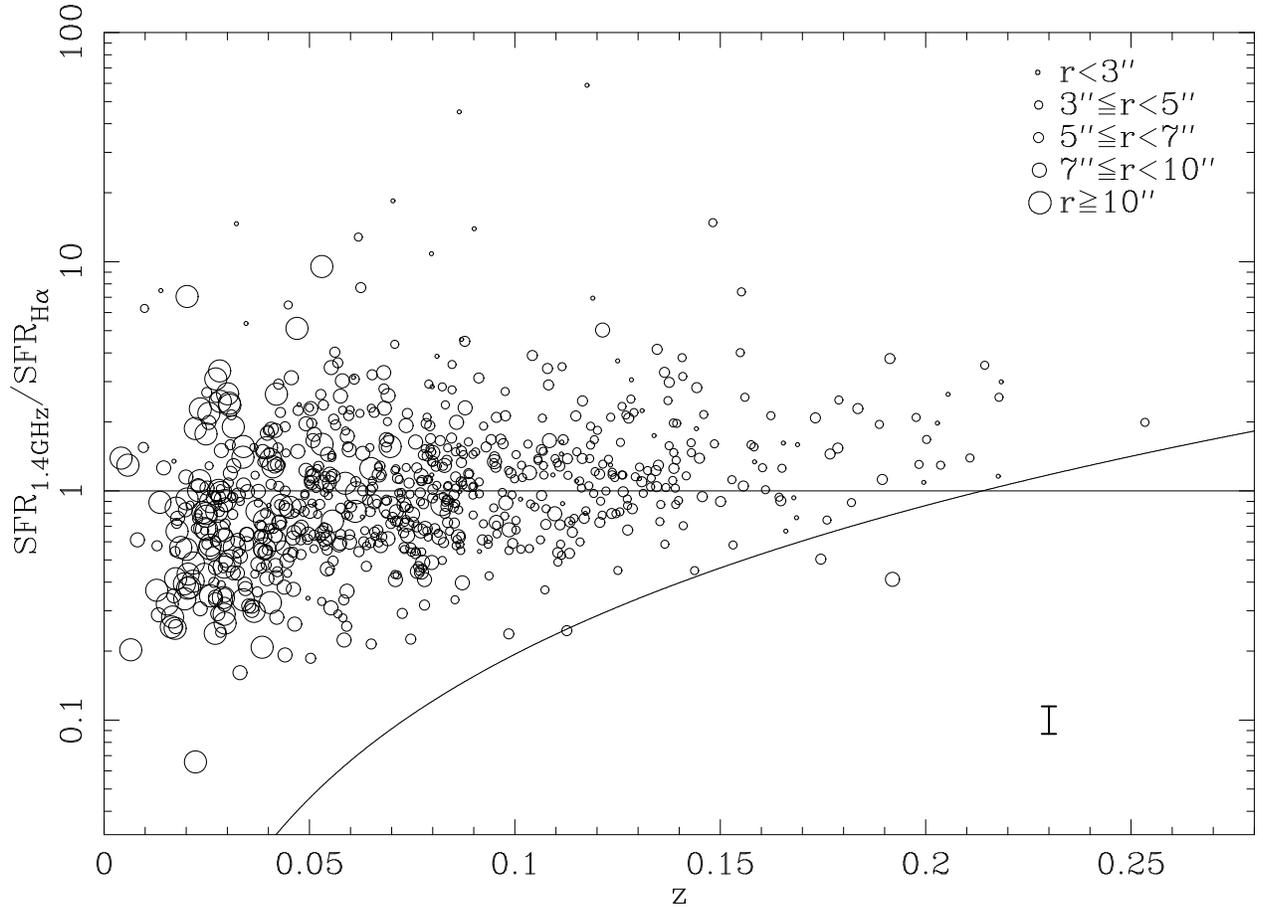}}}
\caption{Ratio of SFRs from 1.4\,GHz and H$\alpha$ luminosities as a
function of redshift. The horizontal line shows the one-to-one relationship,
and the curved line indicates the effect of the flux density limit of the
FIRST survey detections (about 0.75\,mJy) combined with the approximate
upper limit of the measured H$\alpha$ SFRs (about $100\,M_{\odot}\,$yr$^{-1}$).
The symbol sizes reflect the apparent size of the object, based on the
Petrosian radius as indicated in the Figure.
The error bar in the lower right indicates the typical uncertainty in the
SFR ratio. This is again random error from the measurements only.
 \label{sfrdiffs}}
\end{figure}

The differences between the measurements of SFR from H$\alpha$ and 1.4\,GHz
luminosities from Figure~\ref{radhasfr}(c) were further investigated to explore
whether there was any residual trend in the relation. Figure~\ref{sfrdiffs}
shows the ratios of these SFRs as a function of redshift, indicating
the consistency of the two measurements, on average. The effect of the
flux density limit of the radio survey is also shown. This limit
(about 0.75\,mJy, somewhat lower than the completeness level of 2\,mJy) starts
to bias the observed distribution for systems beyond a redshift of
$z\approx0.1$. This bias is in the sense of losing sensitivity to systems with
apparent underestimates of SFR$_{\rm 1.4\,GHz}$ with respect to
SFR$_{\rm H\alpha}$. Also, below a redshift of $z\approx0.03$, the
distribution seems to show predominantly more systems with
SFR$_{\rm H\alpha}$ overestimated compared to SFR$_{\rm 1.4\,GHz}$.
This is unlikely to be a selection effect, and is also unlikely to
be an underestimate in the 1.4\,GHz luminosities, as the NVSS
measurements used for these large, nearby systems are sensitive to the
emission over the full range of sizes seen. This effect is instead more likely
to be an overestimate in the H$\alpha$ luminosities as a result of the
effective aperture correction used. The aperture correction implicitly
assumes that the star formation sampled through the fiber is representative
of the distribution over the whole galaxy, and that this scales directly with
the broadband optical emission. The more realistic scenario is that
(especially in relatively low SFR, nearby systems) the star formation
is patchy, and distributed non-uniformly throughout the galaxy. Hence
it is not surprising that the simple aperture correction used for these
systems produces an overestimate. This overestimate is not seen in
the distribution of measurements between $0.05\lesssim z \lesssim0.1$,
where the 1.4\,GHz flux density limit does not yet affect the observed
distribution. The most likely explanation here is that the lowest redshift
systems have a larger contribution from galaxies with centrally
concentrated star formation, and possibly also from irregular and dwarf
systems, known to contain patchy SF distributions.

It is worth pointing out here that there is a subtle selection effect arising
from the requirement of having a radio detection for the SF sample. This is an
effective emission line flux limit resulting from the radio flux density
limit. If the 1.4\,GHz detection limit of about 0.75\,mJy is transformed to
an H$\alpha$ flux limit via the SFR calibrations of Equations~\ref{sfr1.4} and
\ref{sfrha}, (after incorporating the factor of 20 corresponding to the
average combined obscuration and aperture correction), an effective
H$\alpha$ flux limit of about $3.5\times10^{-18}\,$Wm$^{-2}$ is derived.
This is over an order of magnitude higher than the minimum observed H$\alpha$
fluxes of $\approx3\times10^{-19}\,$Wm$^{-2}$ for the spectroscopically
classified SF galaxies in the whole DR1. A similar relation can be derived
for [O{\sc ii}] line fluxes. In this case the effective [O{\sc ii}] flux limit
($\approx7\times10^{-19}\,$Wm$^{-2}$) imposed by the radio flux density limit
is a little closer to the actual minimum [O{\sc ii}] fluxes observed
($\approx1\times10^{-19}\,$Wm$^{-2}$) in the DR1 SF galaxies. The primary
result of this selection effect seems to be that the radio detected sample is
more similar to a complete sample, in the sense that the incompleteness at low
H$\alpha$ fluxes is removed. This is explored further in \S\,\ref{disc_rad}.

\subsection{{\rm [O{\sc ii}]} Star Formation Rates}
\label{sectiono2}

An SFR calibration based on the [O{\sc ii}] emission is an important tool for
probing galaxy SFRs at $z\gtrsim0.4$ where H$\alpha$ is redshifted out
of the bands easily accessible by optical spectroscopy. As with H$\alpha$,
[O{\sc ii}] luminosities require aperture and obscuration corrections, which
are described in detail in \S\,\ref{hasfr} above. SFRs derived from
[O{\sc ii}] luminosities are based on the fact that there is a good
correlation between observed [O{\sc ii}] line fluxes and observed
H$\alpha$ lines fluxes (i.e., prior to any obscuration corrections).
The SFRs come from using the [O{\sc ii}] emission as a proxy for the
H$\alpha$ emission, which must then be corrected for obscuration based
on the obscuration at the wavelength of H$\alpha$ in order to use
the SFR calibration derived for H$\alpha$ luminosities \citep{Ken:98,Ken:92}.

Many recent estimates of [O{\sc ii}] SFRs rely on the SFR calibration of
\citet{Ken:98}, an average of the calibrations reported by \citet{Gal:89} and
\citet{Ken:92}, each of which are based on samples of fewer than 100 galaxies.
In particular the observed ratio $F_{\rm [OII]}/F_{\rm H\alpha}=0.45$
from \citet{Ken:92} is extensively used. This ratio is, however, luminosity
and metallicity dependent \citep{Jan:01}, and the appropriate ratio
should be determined for a given sample depending on its selection effects.
The current sample of 791 SDSS SF galaxies with radio counterparts includes
752 with measured [O{\sc ii}] emission. Following the methodology used in
\citet{Ken:92}, the ratio of the observed fluxes in the [O{\sc ii}] and
H$\alpha$ emission lines was investigated (see also Figure~\ref{o2hist}, and
subsequent discussion in \S\,\ref{disc_rad}), and the median ratio was found
to be $F_{\rm [OII]}/F_{\rm H\alpha}=0.23$. (This result comes after a stellar
absorption correction EW$_{\rm c}=1.3\,$\AA\ is applied to the H$\alpha$
emission, although the median ratio changes negligibly if this step is
omitted.)

Applying or omitting the aperture corrections when calculating this
ratio also changes the median value negligibly, although it is certainly
true that a differential distribution for the origin of the line
emission would not be accounted for with the aperture correction methods
used here. If [O{\sc ii}] emission came predominantly from galaxy
disks, and H$\alpha$ predominantly from galaxy nuclei, for example,
the current estimate would be skewed low. It seems unlikely, though, that this
contrived geometry of line emission should occur for the majority
of systems. Given, additionally, the expected association of both forms
of emission with star formation regions, it seems reasonable that
both [O{\sc ii}] and H$\alpha$ emission should be at least approximately
colocated throughout SF galaxies.

The median ratio determined for the current sample (from fiber spectroscopy)
is slightly lower than that found by \citet{Jan:01} (using spatially
integrated spectra) for galaxies of similar absolute magnitude.
This does not appear to be an artifact of the radio selected nature of
the current sample. If the radio detection requirement is relaxed, and a
complete, volume-limited sample of optically selected SF galaxies constructed,
an unbiased estimate of the distribution of this flux ratio can be established.
This is pursued further in \S\,\ref{disc_rad} below, and results in median
values for $F_{\rm [OII]}/F_{\rm H\alpha}$ very similar to that found for the
full radio detected sample currently being explored. For the present
discussion we adopt the ratio $F_{\rm [OII]}/F_{\rm H\alpha}=0.23$ as being
representative of typical SF galaxies, given the optical luminosity range
of the present sample. This calibration may also be useful for higher redshift
surveys of galaxies of similar luminosity, where observations of the
H$\alpha$ line are more difficult. The metallicity dependence of the
$F_{\rm [OII]}/F_{\rm H\alpha}$ ratio, however, must also be accounted for
at higher redshifts, given the strong evolution in the metallicity-luminosity
relation \citep{Kob:03}.

Using this estimate for the correlation between [O{\sc ii}] and H$\alpha$
line fluxes the calibration of $L_{\rm H\alpha}$ to SFR
(Equation~\ref{sfrha}) is transformed to
\begin{equation}
\label{o2sfreq}
{\rm SFR_{\rm [OII]}}\,(M_{\odot}\,{\rm yr^{-1}}) =
   \frac{L_{\rm [OII]}}{2.97\times10^{33}\,{\rm W}},
\end{equation}
where $L_{\rm [OII]}$, due to the way this calibration was derived,
must incorporate the obscuration correction valid at the wavelength of
H$\alpha$ \citep{Ken:98}. The final [O{\sc ii}] SFR estimate is explicitly
shown in Appendix~\ref{formulae} as Equation~\ref{o2obssfr}.

The effectiveness of this calibration can be seen in the comparison
of [O{\sc ii}] derived SFRs with those from 1.4\,GHz and H$\alpha$
in Figure~\ref{o2sfr}. The small number of systems in Figure~\ref{o2sfr}(a)
with anomalously high 1.4\,GHz SFRs may be composite AGN/SF systems,
in which the AGN is masked by obscuration at optical wavelengths, since the
H$\alpha$ and [O{\sc ii}] SFRs for these systems are in good agreement.

\subsection{FIR Star Formation Rates}
\label{sectionfir}

FIR flux densities are available from the IRAS catalogs for a sub-sample
of our SF galaxies, and positional matching with the FIRST radio sources
identifies 191 galaxies in this final SF sample with IRAS detections.
The $60\,\mu$m and $100\,\mu$m flux densities, $S_{60}$ and $S_{100}$
can be used to derive a FIR flux following \citet{Hel:88},
FIR(W\,m$^{-2}$)$=1.26\times10^{-14}(2.58\,S_{60}+S_{100})$.
There are several calibrations of FIR luminosity to SFR in the literature,
\citep[for a summary see][]{Ken:98}. We begin with the recent calibration
of \citet{Bell:03}, (from which our chosen 1.4\,GHz SFR calibration
was derived):
\begin{equation}
 \label{belfir}
{\rm SFR_{FIR}}\,(M_{\odot}\,{\rm yr^{-1}}) =
    \frac{1.75 f L_{\rm FIR}}{2.44\times10^{36}\,{\rm W}}
       =\frac{f L_{\rm FIR}}{1.39\times10^{36}\,{\rm W}},
\end{equation}
where
\begin{equation}
\label{firf}
f = 
\left\{
\begin{array}{ll}
 1 + \sqrt{2.186\times10^{35}\,{\rm W}/L_{\rm FIR}} & L_{\rm FIR} > L_c \\
 0.75\,(1 + \sqrt{2.186\times10^{35}\,{\rm W}/L_{\rm FIR}}) & L_{\rm FIR} \le L_c,
\end{array}
\right.
\end{equation}
and $L_c=2.186\times10^{37}\,$W. $L_{\rm FIR}$ is the luminosity
corresponding to the FIR flux as defined above, and the factor of
1.75 in Equation~\ref{belfir} converts this to a luminosity representative
of the full ($8-1000\,\mu$m) mid- to far-infrared spectrum
\citep[for details see][]{Bell:03,Kew:02}. The piecemeal nature of this
Equation comes from the assumption of varying fractions of old stellar
populations with luminosity. It is plausible, however, that galaxies with
high FIR luminosities in the current sample have similar relative
contributions (to FIR luminosity) from old stellar populations as the lower
luminosity systems. This is in contrast to the sample from \citet{Bell:03},
which was more inhomogeneously constructed, possibly resulting in preferential
selection of starbursting galaxies with low old fractions
above $10^{11}\,M_{\odot}$\,yr$^{-1}$ (Bell, 2003, private communication).
Consequently we have chosen to adopt an old stellar population fraction of
30\% independent of luminosity, and apply the portion of Equation~\ref{firf}
valid for $L_{\rm FIR} \le L_c$ to all galaxies. This is explicitly presented
in Equation~\ref{belfir2}. We further recommend this strategy to others working
on samples of radio-selected galaxies (Bell, 2003, private communication).
The resulting SFR$_{\rm FIR}$ is compared with those from 1.4\,GHz and
H$\alpha$ in Figure~\ref{firrad}. This calibration is within a factor
of two of that of \citet{Ken:98} down to about $3\times10^{-3}\,L_c$
\citep{Bell:03}. Note that the empirical relation from \citet{BX:96},
measured from a sample of galaxies of types Sb and later, produces SFRs almost
80\% larger than this calibration, and that of \citet{Con:92} produces SFRs
about $60\%$ larger.

\subsection{$u$-Band Star Formation Rates}
\label{sectionsfru}

While ultraviolet (UV) luminosity ($\lambda\lesssim2500\,$\AA) is
commonly used as an SFR indicator, the luminosity at $u$-band wavelengths
($\lambda\approx3600\,$\AA) is similarly dominated in starburst galaxies
by young stellar populations, and in the absence of UV measurements the
$u$-band luminosity may thus be used instead as an SFR indicator
\citep{Cram:98}. The $u$-band luminosities used here are derived from
the SDSS k-corrected absolute $u$-band magnitudes \citep{Bla:03}, after
incorporating an obscuration correction based on the Balmer decrement and
the extinction curve from \citet{Cal:00}. It is worthwhile pointing out
that the average $u$-band obscuration correction ranges from a factor of 3
at SFRs of $1\,M_{\odot}\,$yr$^{-1}$ up to about a factor of 10 at SFRs
of $100\,M_{\odot}\,$yr$^{-1}$.

UV luminosities have been extensively used as SFR indicators,
and for wavelengths $1500\,$\AA\,$\lesssim\lambda\lesssim2500\,$\AA\ the
calibration given by \citet{Ken:98}
\begin{equation}
{\rm SFR_{UV}}\,(M_{\odot}\,{\rm yr^{-1}}) =
   \frac{L_{\rm UV}}{7.14\times10^{20}\,{\rm W\,Hz^{-1}}}
\end{equation}
has proven quite effective. For $u$-band luminosities, though, it is somewhat
more difficult to assign a simple scaling factor to derive an SFR due to
the strong dependence on the evolutionary timescale. From synthetic galaxy
spectra \citep[e.g.,][]{FR:97} it can be seen that the $u$-band luminosity
varies from about factor of 10 lower than the UV luminosity at the onset of
a burst of star formation, to almost equivalent by $10^8$ years later
\citep[see also discussions of the dependence of UV measures on the
timescale of SFR in][]{Sul:00,Gla:99}.

Given this sensitivity of the $u$-band (and indeed UV) luminosity to the
starburst age and the assumed star formation history, a more complex
calibration is in general likely to be necessary. This may take the form
of a non-linear dependency on $L_{\rm U}$ to reflect the rapid change of
$L_{\rm U}$ with respect to $L_{\rm UV}$ during the first $10^8$ years of a
starburst, and to account for the presence of old stellar populations that are
likely to contribute significantly to $L_{\rm U}$ in less luminous systems
\citep{Bell:03}.
More quiescent or low-luminosity SF systems have a relatively larger
contribution to their $u$-band luminosity from old stellar populations, (they
are, on average, redder than more luminous galaxies), and this causes a linear
calibration from luminosity to SFR to result in an overestimate of the SFR.
While these effects can be modelled in detail using stellar spectral synthesis
methods \citep{Sul:00}, for the purposes of a simple SFR estimate based on
a measured luminosity an empirical non-unity power-law relationship between
$u$-band luminosity and SFR can be derived. Although providing no information
about the SFR history or the SED evolution, such a calibration has the
advantage of being easy to construct and can form a useful tool in
subsequent analysis.

To construct such a tool we require a complete, volume-limited sample of
galaxies, a necessity not demanded of the previous sections where
straightforward comparisons between existing calibrations were being explored.
This requirement is necessary to avoid biasing the calibration low or high
based on the independent limits of the H$\alpha$ and $u$-band detections.
In order to have the largest complete sample from which to derive the new
calibration, and to eliminate any potential concerns introduced regarding
radio selection, we thus construct an SF galaxy sample from the DR1 based on
the completeness criteria described by \cite{Gom:03}, and do not require
radio detection at all. The criteria used specify that galaxies should lie
closer than $z\leq0.095$ (since aperture corrections are applied, no lower
redshift limit is necessary) and have $M_{\rm r}\leq-20.57$ (after converting
to our chosen value of $H_0$). The small number of systems with $z<0.05$
\citep[the lower redshift limit used by][]{Gom:03} which show slight
overestimates in their aperture corrected SFR$_{\rm H\alpha}$ relative to
SFR$_{\rm 1.4GHz}$ do not bias the result, which remains unchanged if these
objects are excluded. A further restriction was applied, a limit on
the H$\alpha$ fluxes, $F_{\rm H\alpha}>3\times10^{-18}\,$Wm$^{-2}$. This
corresponds to an SFR of about $0.5\,M_{\odot}$yr$^{-1}$ at the redshift
limit of the complete sample, and ensures no bias will be introduced to
the derived calibration through the presence of incompletely sampled fainter
H$\alpha$ systems. This results in a sample of 2625 spectroscopically
classified SF galaxies with measured H$\alpha$ and $u$-band luminosities.

Using this sample the ordinary least-squares bisector method of linear
regression \citep{Iso:90} was applied to $\log(L_{\rm U})$ (after applying
obscuration corrections based on the Balmer decrements) and
$\log({\rm SFR_{H\alpha}})$. It is recognised that the luminosity limits
imposed by requiring completeness will cause a small bias to this fit, (since
the imposed limit on $r$-band luminosity translates to an effective limit
on $u$-band luminosity, and omitting galaxies with lower luminosities affects
the slope of the fit slightly), nevertheless this still produces a better
estimate of the true relation than were an incomplete sample used. The
comparison for the complete sample of galaxies between $L_{\rm U}$ and
SFR$_{\rm H\alpha}$ can be seen in Figure~\ref{allhausfr}(a), which indicates
the resulting best fit relation. This relation results in the calibration:
\begin{equation}
{\rm SFR_U}\,(M_{\odot}\,{\rm yr^{-1}}) = \left(\frac{L_{\rm U}}
  {1.81\times10^{21}\,{\rm W\,Hz^{-1}}}\right)^{1.186},
\end{equation}
and the result of applying this calibration is shown
in Figure~\ref{allhausfr}(b). Strictly speaking, this calibration is only
valid over the range of luminosities probed here,
$2\times10^{21}\lesssim L_U \lesssim 10^{23}\,$W\,Hz$^{-1}$.
The explicit calculation in terms of the observable quantities is given in
Equation~\ref{usfr}, in Appendix~\ref{formulae}.
Since the calibration from luminosity
to SFR is no longer linear, a multiplicative obscuration correction cannot
be applied directly to the SFR in the same manner as to the luminosity.
Rather the appropriate exponent as given in the SFR calibration must be
incorporated as well.

The entire sample of 21649 SF galaxies with measured $u$-band and
H$\alpha$ SFRs are compared in Figure~\ref{allhausfr2}. The effects of the
incompleteness can be seen, appearing as a slight apparent bias to
overestimated $u$-band SFRs. The fitted relation, therefore, is sensitive to
the completeness of the sample being used, emphasising the need
for the initial calibration to be performed on a well-defined sample.

Using the new SFR calibration for $u$-band luminosity, SFR$_{\rm U}$ is
shown as a function of SFR$_{\rm 1.4\,GHz}$ in Figure~\ref{firstusfr}.
The rms deviation between SFR$_{\rm U}$ and SFR$_{\rm 1.4\,GHz}$ is
0.23 dex, or a factor of 1.7 either side of the one-to-one line.

\section{Discussion}
\label{disc}

\subsection{Systematics and absolute SFR calibrations}
\label{disc_abs}

We have demonstrated a consistency between independent SFRs estimated from
H$\alpha$ and 1.4\,GHz luminosities, used an observed H$\alpha$/[O{\sc ii}]
flux correlation to establish the consistency of the [O{\sc ii}] SFRs,
identified the FIR calibration giving most consistent SFRs, and derived
a $u$-band calibration defined to give consistent SFRs. It should be
emphasised that, since the calibration for SFR$_{\rm 1.4\,GHz}$ is derived
from the FIR calibration \citep{Bell:03}, and since the [O{\sc ii}] and
$u$-band calibrations are derived from the H$\alpha$ calibration, the
only independent calibrations being directly compared are those of
H$\alpha$ and FIR, both ultimately coming from \citet{Ken:98}. This
is still non-trivial, however, given the different physics involved
in the two derivations (ionising luminosities compared with total bolometric
fluxes), and the broad consistency seen across the range of wavelengths
explored is highly encouraging. This consistency is in spite of the large
scatter, frequently referred to above, in the SFR values measured at different
wavelengths for individual systems. The dispersions about the one-to-one line
for each estimator compared with the 1.4\,GHz estimator are summarised in
Table~\ref{sfrtab}, where the uncertainties shown are the rms deviations
either side of the one-to-one line. Some issues that are likely to contribute
to this observed scatter include (1) the reliability of the assumptions
underlying the aperture corrections to the emission line estimates;
(2) whether the Balmer decrement obscuration estimate (which is measured only
for the region seen through the fiber aperture) is valid for the whole galaxy;
(3) the extent to which any low luminosity AGN that may be present in the SF
galaxies contributes to the observed radio luminosity; (4) galaxy to galaxy
differences in the average 1.4\,GHz luminosity generated by supernovae, either
through differences in electron population densities, temperatures, or
confinement, strength of interstellar magnetic fields, or other physical
differences.

It is possible that in addition to this observed scatter, however,
the {\em absolute\/} SFR calibrations are still uncertain by up to a factor of
two \citep{Bell:03,Con:02}. Apart from concerns arising through the detailed
physics involved in deriving the SFR calibrations, the only
additional systematics that might change the quantitative SFRs derived
herein are the assumed equivalent width correction to account for
stellar absorption, and the chosen obscuration curve. The aperture corrections
could produce underestimates of SFR$_{\rm H\alpha}$ only for galaxies large
enough that the fiber predominantly samples light from the nucleus, the star
formation occurs primarily in the disk, and the galaxy is observed close
to face on. Hence it seems likely that any bias caused by the aperture
corrections will be overestimates of SFR$_{\rm H\alpha}$, as seen in
Figure~\ref{sfrdiffs}, and that these are relatively small effects restricted
to very nearby systems. (It is of course possible that the assumptions
involved in making the aperture correction contribute, perhaps significantly,
to the scatter seen in the measured individual SFRs.)

Differences between estimates of the obscuration curve appropriate for
correcting emission lines produce only small changes (of order $5\%$) from the
results given here. The assumption of Case B recombination may also contribute
to the scatter between the different SFR estimates, since this will not
be valid for all systems (although it should be reasonable for the
majority of the SF galaxies). It is also possible that the detailed geometries
of gas and dust in individual objects could result in quite different
obscurations than estimated here, but for this sample of relatively low
redshift galaxies the obscuration curve adopted should be fairly
representative. These effects, while contributing to the observed scatter,
should not act in a systematic fashion.

The H$\alpha$ SFR values reported here could potentially be increased
by up to about $20\%$ by reducing the chosen EW correction for stellar
absorption (a value of ${\rm EW_c}=0.7\,$\AA\ increases the derived SFRs by
this factor). This arises through the change introduced in the Balmer
decrement when a different value of the ${\rm EW_c}$ is used. Since the
measured H$\beta$ emission is in many cases comparable to the correction
being applied to it, the resulting Balmer decrement is very sensitive to the
chosen value. Values of ${\rm EW_c}$ smaller than about 0.7\,\AA\ would,
however, no longer be consistent with the stellar absorption corrected
Balmer decrements derived from the flux-summing method of line measurement.
Making the stellar absorption correction smaller also increases the
discrepancy between the two methods (described in Appendix~\ref{apcor2}) of
estimating SFR$_{\rm H\alpha}$. So there is a limit of about $20\%$ by
which SFR$_{\rm H\alpha}$ can be increased through the chosen estimate of
${\rm EW_c}$. Even in combination with the other systematic uncertainties in
the H$\alpha$ SFR calibration, the values reported here seem reliable in an
absolute sense to better than a factor of two. In any case, it doesn't seem to
be possible to produce high enough values of SFR$_{\rm H\alpha}$ to become
consistent with the 1.4\,GHz SFR calibration of \citet{Con:92} (which gives
SFRs about a factor of two higher than the calibration used here), without
revising the H$\alpha$ SFR calibration.

Since the SFR$_{\rm [OII]}$ calculation uses the same obscuration correction
as for H$\alpha$, and since the $u$-band SFRs are calibrated to
SFR$_{\rm H\alpha}$, these would both be changed consistently as different
estimates of ${\rm EW_c}$ change SFR$_{\rm H\alpha}$ in the above case.
The FIR estimate, similarly, is consistent with the chosen 1.4\,GHz
estimate (by construction), so the only truly independent SFR estimators
in the current investigation are SFR$_{\rm H\alpha}$ and SFR$_{\rm 1.4GHz}$.
The consistency between these two results does, however, support the
reliability of the assumed stellar absorption correction, and since the
absolute calibration of the H$\alpha$ SFRs appears quantitatively reliable,
to better than a factor of two, the radio and FIR calibrations adopted
must be similarly reliable. This conclusion regarding the quantitative
reliability of the SFR calibrations, of course, only applies to the overall
population. Measurements of individual objects, as can be seen from the large
scatter in all the SFR comparison diagrams, can still easily be uncertain by
factors of two or more.

So, despite the discrepancies (factors of two) in many of the available
calibrations, a strong argument can be made that the absolute SFR calibrations
should be close to those adopted here. This preliminary exploration suggests
that the presently adopted calibrations give SFRs which are on average
reliable to better than a factor of two.

\subsection{Radio-selected SF galaxies}
\label{disc_rad}

The properties of the radio detected SF galaxies have been explored
to investigate some details of how they differ from optically selected
SF galaxies. The complete sample of SF galaxies constructed
for the $u$-band calibration in \S\,\ref{sectionsfru}, above, was used
to compare the distribution of various measured parameters between
the optically selected and radio detected objects. There are 2625
galaxies in the complete sample, of which 380 are radio detected.
Figure~\ref{maghist} indicates the distribution in optical luminosity
spanned by both the complete sample and the radio detected galaxies
within that sample. The median magnitudes of these two distributions
are similar ($-20.6$ for the complete sample, $-20.8$ for the radio
detected systems), although the radio detected systems show a much
broader magnitude distribution. Gaussian fits to the two histograms
indicate a full width at half maximum (FWHM) of 0.9\,mag for the
complete sample, and 1.3\,mag for the radio detected galaxies.
Figure~\ref{allvsrad} shows the distributions of rest frame $u-g$ and
$u-r$ colors, concentration index inverse, $1/c$, and D$_{4000}$ for
both the complete SF sample and the radio detected galaxies in
the complete sample. (Concentration index, here, is defined to be
the ratio of the Petrosian radius enclosing $90\%$ of a galaxy's light
to that enclosing $50\%$, e.g., \citeauthor{Kau:03} \citeyear{Kau:03}.)
It is clear that the radio detected galaxies
preferentially identify a sub-population of the SF galaxies, with
somewhat redder colors, more concentrated morphologies, and higher
D$_{4000}$ values. This all suggests that the radio selected SF
galaxies favour systems with a greater relative contribution from
old stellar populations than optically selected systems. It should be noted
here that while the radio detected SF systems are redder and bulgier
than optically selected SF galaxies, they are still on average
bluer and diskier than the overall population of galaxies. This is confirmed
by the dashed histograms in Figure~\ref{allvsrad}, which show the distribution
of parameters for a complete sample of all galaxies, constructed in an
identical fashion to the complete sample constructed above, but without any
restriction on the H$\alpha$ EW. This new complete sample of 24444 galaxies
clearly shows bimodal distributions corresponding to red and blue galaxy
populations, and verifies that the radio-detected SF galaxies
truly represent a sub-population of SF galaxies, and are not the
result of an unexpected selection effect, or some unexplained early-type
galaxy contamination to the SF sample.

The broader distribution seen in absolute magnitude for the radio
detected galaxies suggests that these results may simply reflect a
tendency for the radio detection to favour larger, brighter galaxies.
But the distributions in the $u$-band luminosities, for the radio detected
galaxies compared with the complete sample, suggests that this may not
be the whole picture. In Figure~\ref{uvshacplt} the $u$-band and
H$\alpha$ SFRs are compared for the galaxies in the complete sample, as for
Figure~\ref{allhausfr}(b), but now identifying the radio detected galaxies.
There seems to be a preference for the radio detected systems to appear below
the one-to-one line, at least for low to moderate SFRs. Since the comparison
between radio and H$\alpha$ SFRs shows no such effect, it seems likely that
the radio detection preferentially identifies {\em lower\/} $u$-band
luminosity systems for a given SFR.

The reasons why these types of preferential selection occur may be related to
the optically selected samples undersampling the red end of the distribution
as a result of obscuration effects, but it is also possible that the
radio detection may undersample the blue end of the distribution due to
the different physical processes and timescales producing the emission
in the different wavelength regimes. A full exploration of the questions
raised by these and related results is clearly warranted, although, being
beyond the scope of the present work, this is being investigated in detail in
a subsequent paper \citep{Hop:03}.

The radio detected systems seem to have a notably higher median
obscuration than the complete sample, $A_{\rm H\alpha}=1.6\,$mag compared
with $A_{\rm H\alpha}=1.2\,$mag (Figure~\ref{bdhist}), and the distribution
is broader, a Gaussian fit giving a FWHM of 1.2\,mag compared
with 1.0\,mag for the optically selected galaxies. The median value here
for the optically selected objects is consistent with the 1.1\,mag
of obscuration commonly assumed for H$\alpha$ measurements of SF
galaxies \citep{Ken:83}. The median value for the radio detected systems,
however, is significantly higher, and may be related to the efficiency of
radio measurements in detecting heavily obscured sources. In other words,
a radio selected sample should not bias against galaxies with high obscuration,
producing a distribution of obscurations characteristic of the true
distribution \citep{Afo:03}. The slightly higher median and broader
distribution for the radio detected systems suggests that it may be likely
that the optically selected sample does indeed undersample the redder,
bulgier end of the galaxy distribution. This does not, however, exclude the
possibility that radio detection simultaneously undersamples the bluer end,
as suggested by the lack of higher SFR$_{\rm U}$ systems seen in
Figure~\ref{uvshacplt}.

In Figure~\ref{o2hist}(a), histograms of the $F_{\rm [OII]}/F_{\rm H\alpha}$
flux ratio for the (incomplete) sample of all the SF galaxies from DR1
shows a median value of 0.38, similar to that found by \citet{Ken:92}.
The radio selected systems here have a much lower ratio, 0.23, more
comparable to the median values found for the complete samples, shown in
Figure~\ref{o2hist}(b), for both optically selected and radio detected
galaxies (0.26 and 0.21 respectively).
It appears, moreover, that there is a weak trend between the ratio
$F_{\rm [OII]}/F_{\rm H\alpha}$ and H$\alpha$ line flux or EW, the latter
being shown in Figure~\ref{fluxrat2}, in the sense of higher flux ratios
in higher EW systems. This is likely to be directly related to the correlation
with luminosity noted by \cite{Jan:01}, as the higher EW systems include those
with the lower luminosities. The discrepancy between the present result and
those from \citet{Gal:89} and \citet{Ken:92} can also be explained by
some level of incompleteness in those samples at the low EW end. This would
lead to a bias toward higher EW systems, and hence higher
$F_{\rm [OII]}/F_{\rm H\alpha}$ ratios. For systems in the
complete sample with EW(H$\alpha$)$>70\,$\AA\ the mean ratio is 0.46,
much closer to the mean of 0.45 found by \citet{Ken:92}.

An important conclusion here is that the use of [O{\sc ii}] luminosities as
an SFR estimator requires a good understanding of the selection effects
of the sample being investigated. The notion that SFRs based on [O{\sc ii}]
luminosities are not very reliable \citep{Jan:01,CL:01,Ken:98} is perhaps
related at least partially to the incomplete understanding of the
$F_{\rm [OII]}/F_{\rm H\alpha}$ distribution valid for the sample under
consideration, and what governs this distribution for the given set of
selection effects. To the extent that the $F_{\rm [OII]}/F_{\rm H\alpha}$
ratio depends on metallicity, for example, the evolution of the
luminosity-metallicity relation \citep{Kob:03} must also be taken into account
for higher redshift systems.

A brief exploration of how $F_{\rm [OII]}/F_{\rm H\alpha}$ varies with several
parameters was pursued, to identify any obvious trends that might help in
this type of sample characterisation. \citet{Jan:01} find a strong
correlation between $F_{\rm [OII]}/F_{\rm H\alpha}$ and $M_B$, with
brighter galaxies having lower flux ratios. No such relation (with $M_g$,
$M_r$ or $M_z$) is convincingly measurable in the present sample although
this only spans a range of about three magnitudes, much smaller than the
eight magnitudes spanned by the sample of \citet{Jan:01}. Given the scatter
in the relation shown in Figure~1 of \citet{Jan:01}, a range of at least
5 or 6 magnitudes would need to be sampled to identify this trend, so it is
not surprising that it is not clearly detected in the current sample.
There is, however, a suggestion in the current data that the brightest
systems are restricted to lower ratios, and this weak trend is reflected
in the weak trend mentioned above with H$\alpha$ EW. There is, similarly,
very little detected trend with galaxy size, although again the very largest
galaxies are restricted to lower ratios.

The connection between the $F_{\rm [OII]}/F_{\rm H\alpha}$ ratio and the
Balmer decrement was also examined. If a common obscuration curve
is valid for all the galaxies in the sample, there should be a simple
relationship between this ratio and the $F_{\rm H\alpha}/F_{\rm H\beta}$
ratio. Figure~\ref{o2bd} shows this relation for the complete sample
of SF galaxies \citep[compare with Figure~2(c) of][]{Jan:01}.
The curves shown in Figure~\ref{o2bd} indicate the relationship
expected for the obscuration curve of \citet{Car:89} and intrinsic
$F_{\rm [OII]}/F_{\rm H\alpha}$ flux ratios (before any obscuration affects
the emission lines) of 0.3, 1.0 and 2.0. This suggests either that the
complete sample displays a range of these intrinsic ratios spanning about
$0.3-2.0$, or that the attenuation in SF galaxies does not follow a unique
attenuation law owing to varying dust geometries, changes in dust properties,
or both.

Another comment can be made regarding Figures~\ref{fluxrat2} and \ref{o2bd}.
A higher value for $F_{\rm [OII]}/F_{\rm H\alpha}$ corresponds
to lower obscuration (from Figure~\ref{o2bd}), and EW(H$\alpha$) is
a tracer of the current relative SF (in the sense of absolute current SFR
relative to the total integrated SFR). So the weak trend seen in
Figure~\ref{fluxrat2} might be taken to suggest that systems with
higher {\em relative\/} SFRs show (on average) {\em lower\/} obscurations.
This is an intriguing result given that Figure~\ref{bdecvssfr} implies
systems with higher {\em absolute\/} SFRs have (on average) {\em higher\/}
obscurations. These results may be useful in further exploring the
nature of obscuration in SF galaxies as a function of their physical
properties. One hypothesis which will be examined in more detail in a
subsequent investigation is that these results are both consistent with
obscuration being primarily a function of galaxy mass or luminosity (larger
galaxies having higher obscurations).

\section{Summary}
\label{summ}

From a sample of 3079 SDSS galaxies having radio luminosities
from the FIRST survey, we have used optical spectroscopic diagnostics
to identify a sub-sample of 791 SF galaxies. Using this
sub-sample we have investigated five SFR indicators based on
H$\alpha$, [O{\sc ii}], $u$-band, FIR and 1.4\,GHz luminosities. The
FIRST 1.4\,GHz derived SFRs below about $10\,M_{\odot}\,$yr$^{-1}$ are
progressively underestimated as the galaxy sizes become larger, consistent
with a known limitation of the FIRST survey. This can be corrected by
using NVSS measurements where available for the larger systems.
After applying appropriate obscuration
and aperture corrections, the H$\alpha$ SFR estimate is seen to be consistent
with the 1.4\,GHz estimate, although a large scatter still remains.
The median [O{\sc ii}] to H$\alpha$ flux ratio is found to be
$F_{\rm [OII]}/F_{\rm H\alpha}=0.23$, about a factor of two lower than
commonly assumed, and an updated SFR calibration for [O{\sc ii}] luminosities
was derived to account for this. The resulting [O{\sc ii}] SFRs are highly
consistent with those from H$\alpha$ and 1.4\,GHz luminosities.
A power-law calibration between $u$-band luminosities and SFR$_{\rm H\alpha}$
was empirically derived, which provides measurements of SFR based on
$L_{\rm U}$ consistent with the other estimators, although more detailed
investigation of the physical processes driving this relation are warranted.
Issues surrounding the reliability of the absolute SFR calibration were
addressed, suggesting that the calibrations used here should be
reliable to better than a factor of two.

With these results there are now three reliable SFR estimators
available from the SDSS measurements, H$\alpha$, [O{\sc ii}], and
$u$-band luminosities. With the large sample size and the extensive
spectroscopic measurements archived by the survey, it thus provides a
vast resource for investigations of star formation in the universe.

Investigating the properties of the SF galaxies, it is found that the
median obscuration at the wavelength of H$\alpha$ for the complete sample of
SF galaxies is A$_{\rm H\alpha}=1.2\,$mag, comparable with the 1.1\,mag
from \citet{Ken:83}, while the radio detected systems are
notably higher, A$_{\rm H\alpha}=1.6\,$mag. The properties of the
radio detected sample imply that radio selection preferentially identifies
somewhat redder, bulgier SF systems, (although still bluer and diskier than
the general galaxy population), having a relatively larger contribution
from the old stellar population, than seen in optically selected SF samples.
This is characterised through redder optical colors, and higher D$_{4000}$
values and concentration indices than in optically selected samples.
This could be attributed to either or both of the cases that
(1) the optically selected samples undersample the red end of the
distribution due to obscuration effects, and (2) radio detection undersamples
the blue end of the distribution due to the different physical processes and
timescales producing the multiwavelength emission.

\acknowledgements

The authors would like to thank the referee, Eric Bell, for many constructive
comments that improved this paper. We also thank Jose Afonso, Ivan Baldry,
Jarle Brinchmann, Daniela Calzetti, Lisa Kewley, Ravi Sheth, Ian Smail, Mark
Sullivan, Regina Schulte-Ladbeck and Dan Vanden Berk for helpful discussion.
AMH acknowledges support provided by the National Aeronautics and Space
Administration (NASA) through Hubble Fellowship grant
HST-HF-01140.01-A awarded by the Space Telescope Science Institute (STScI).
AMH and AJC acknowledge support provided by NASA through grant numbers
GO-07871.02-96A and NRA-98-03-LTSA-039 from STScI, and AISR grant NAG-5-9399.
STScI is operated by the Association of Universities for Research in
Astronomy, Inc., under NASA contract NAS5-26555.

Funding for the creation and distribution of the SDSS Archive has been
provided by the Alfred P. Sloan Foundation, the Participating Institutions,
NASA, the National Science Foundation, the U.S. Department of Energy,
the Japanese Monbukagakusho, and the Max Planck Society.
The SDSS Web site is http://www.sdss.org/.

The SDSS is managed by the Astrophysical Research Consortium (ARC) for
the Participating Institutions. The Participating Institutions are the
University of Chicago, Fermilab, the Institute for Advanced Study, the
Japan Participation Group, The Johns Hopkins University, Los Alamos
National Laboratory, the Max-Planck-Institute for Astronomy (MPIA),
the Max-Planck-Institute for Astrophysics (MPA), New Mexico State
University, the University of Pittsburgh, Princeton University, the
United States Naval Observatory, and the University of Washington.

\begin{appendix}
\section{Aperture corrections}
\label{apcor2}

In addition to the aperture correction described by Equation~\ref{ewsfr},
the H$\alpha$ luminosity calculated using the line flux can be
explicitly aperture corrected. This is done using the ratio of the fluxes
corresponding to the total galaxy magnitude, and the magnitude ``through
the fiber," where this latter term is also one of the outputs of the
photometric pipeline. This fiber magnitude comes from a photometric
measurement of the magnitude in an aperture the size of the fiber, and
is corrected for seeing effects. The Petrosian magnitude was used to
represent the total galaxy flux. The extent of this aperture correction
can be expressed as
\begin{equation}
\label{apeq}
A = 10^{-0.4(r_{\rm Petro} - r_{\rm fiber})}
\end{equation}
where $r_{\rm fiber}$ is the $r$-band fiber magnitude.
The explicit calculation of the aperture corrected luminosity using this
method is thus
\begin{equation}
\label{apsfr}
L_{\rm H\alpha}\,({\rm W}) = 4\pi D_l^{2}\,S_{\rm H\alpha}\,
                 10^{-0.4(r_{\rm Petro} - r_{\rm fiber})},
\end{equation}
with $S_{\rm H\alpha}$ being the stellar absorption corrected flux of
the H$\alpha$ emission, and $D_l$ the luminosity distance. The obscuration
correction has not been included in the above Equation.

Both methods for estimating an aperture-corrected emission line luminosity
give consistent results, as can be seen for H$\alpha$ in
Figure~\ref{hasfrcomp} (an almost identical comparison results for
[O{\sc ii}]). This indicates the high level of self-consistency between the
photometric and spectroscopic data in the SDSS, and suggests that both the
measurements and the methods are self-consistent. There remains a small
systematic discrepancy, though, such that SFRs calculated using the aperture
corrections based on Equation~\ref{apeq} are about 15\% larger. This is
the case for both H$\alpha$ and [O{\sc ii}], and may be related to the
spectrophotometric calibration. The SDSS takes spectra during
conditions which are not deemed ``photometric". The seeing conditions are
thus typically worse for the spectroscopic data than for the photometric data.
Early versions of the photometric pipeline software, used to photometrically
calibrate the spectra, did not take these seeing differences into account.
In the DR1, however, all photometric data used in calibrating the spectra are
convolved to $2''$ seeing (typical of the seeing conditions for spectroscopic
observations). The currently available spectra have, unfortunately, not yet
been recalibrated, resulting in spectrophotometric magnitudes fractionally
brighter than what one would expect from the photometrically measured
fiber magnitudes (by $\approx 0.1$ magnitude in the mean). The SFRs using
the aperture correction of Equation~\ref{apsfr} and the emission line fluxes
(Equation~\ref{apobssfr2}) thus slightly overestimates the correction (since
the line fluxes are slightly overestimated). The alternative SFR estimate
(Equation~\ref{apobssfr}), which uses the line EWs and absolute magnitudes,
is insensitive to this issue.

We investigated the aperture corrections of Equation~\ref{apeq} using Petrosian
and fiber magnitudes in both $r$ and $z$. The $r$ magnitudes were
seen to give a distribution with less scatter than the $z$ magnitudes,
and as a result we chose to apply those in preference. For the estimation
of [O{\sc ii}] luminosities, the $u$-band Petrosian and fiber magnitudes were
used in applying Equation~\ref{apeq}, but very little difference results
if the $r$-band magnitudes are retained instead.

To emphasise the extent of the aperture corrections we show how they vary with
several parameters. The logarithm of the aperture corrections is
shown as a function of galaxy size (the Petrosian radius) in
Figure~\ref{apcorr}, and Figure~\ref{apcorz} shows the aperture correction
as a function of redshift and of SFR$_{\rm 1.4\,GHz}$. It can be seen from
these Figures that the aperture corrections are typically at least a factor
of two and can be as much as an order of magnitude. Finally, the variation
in the ratio of the SFRs from H$\alpha$ and 1.4\,GHz luminosities with the
aperture correction is shown in Figure~\ref{apcorsfrs}. While the relation
is almost flat, as would be desired for an aperture correction introducing
no bias in the derived SFR, there is a measurable positive slope, which
reflects the implicit assumption of a uniform SF distribution throughout each
galaxy. In systems with the largest aperture corrections, the
SFR$_{\rm H\alpha}$ is slightly overestimated, as already seen in
Figure~\ref{sfrdiffs}. There are a small number of systems with very small
aperture corrections, and another small population at all values of aperture
correction, where SFR$_{\rm 1.4GHz}$ appears to be strongly overestimated,
possibly due to the presence of low luminosity AGN components contributing
to the radio emission. Apart from these systems, at low aperture corrections
the median ratio of the two SFRs is unity.

\section{SDSS SFR formulae}
\label{formulae}
For ease of reference, the formulae for deriving SFRs using the measured
SDSS parameters, and the SFR calibrations used, are all collected together
here. The sections in which each formula is derived are also given.
All SFRs are calibrated based on a Salpeter IMF and a mass range of
$0.1<M_{\odot}<100$.

\noindent The H$\alpha$ luminosity to SFR calibration used is
\begin{equation}
{\rm SFR_{H\alpha}}\,(M_{\odot}\,{\rm yr^{-1}}) =
   \frac{L_{\rm H\alpha}}{1.27\times10^{34}\,{\rm W}}.
\end{equation}
For H$\alpha$ SFRs, using the aperture correction method of
Equation~\ref{ewsfr}, the derivation in \S\,\ref{apcor} gives
\begin{eqnarray}
\label{apobssfr}
{\rm SFR}_{\rm H\alpha}\,(M_{\odot}\,{\rm yr^{-1}}) & = &
  ({\rm EW(H\alpha)+EW_c})\,10^{-0.4(M_{\rm r}-34.10)} \nonumber\\
  & & \times \frac{3\times10^{18}}{(6564.61(1+z))^2}
  \left(\frac{S_{\rm H\alpha}/S_{\rm H\beta}}{2.86}\right)^{2.114}
  \frac{1}{1.27\times10^{34}},
\end{eqnarray}
where $S_{\rm H\alpha}$ and $S_{\rm H\beta}$ are the stellar absorption
corrected line fluxes, calculated as in Equation~\ref{stelabs}.
The exponent on the Balmer decrement term (in all the equations given here)
is equal to $k(\lambda)/[k({\rm H\beta})-k({\rm H\alpha})]$, and depends on
the assumed obscuration curve. For obscuration corrections to emission line
luminosities, we assume the obscuration curve of \citet{Car:89} as
recommended by \citet{Cal:01}. ${\rm EW_c}=1.3\,$\AA\ is a reasonable
approximation for the stellar absorption correction when using the SDSS
pipeline spectral line measurements, and corresponds roughly to a
$2.6\,$\AA\ EW stellar absorption in the SF galaxies. Using the alternative
aperture correction given in Appendix~\ref{apcor2} results in
\begin{equation}
\label{apobssfr2}
{\rm SFR}_{\rm H\alpha}\,(M_{\odot}\,{\rm yr^{-1}}) =
  4\pi D_l^{2}\,S_{\rm H\alpha}\,10^{-0.4(r_{\rm Petro} - r_{\rm fiber})}
  \left(\frac{S_{\rm H\alpha}/S_{\rm H\beta}}{2.86}\right)^{2.114}
  \frac{1}{1.27\times10^{34}},
\end{equation}
where $D_l$ is the luminosity distance, and $S_{\rm H\alpha}$ is the stellar
absorption corrected H$\alpha$ line flux.

\noindent The [O{\sc ii}] luminosity to SFR calibration used is
\begin{equation}
\label{o2sfreq2}
{\rm SFR_{\rm [OII]}}\,(M_{\odot}\,{\rm yr^{-1}}) =
   \frac{L_{\rm [OII]}}{2.97\times10^{33}\,{\rm W}},
\end{equation}
where $L_{\rm [OII]}$ incorporates the obscuration correction valid for
the wavelength of H$\alpha$.
For [O{\sc ii}] SFRs the derivation of \S\,\ref{sectiono2} gives
\begin{eqnarray}
\label{o2obssfr}
{\rm SFR}_{\rm [OII]}\,(M_{\odot}\,{\rm yr^{-1}}) & = &
  {\rm EW(OII)}\,10^{-0.4(M_{\rm u}-34.10)} \nonumber\\
  & & \times \frac{3\times10^{18}}{(3728.30(1+z))^2}
  \left(\frac{S_{\rm H\alpha}/S_{\rm H\beta}}{2.86}\right)^{2.114}
  \frac{1}{2.97\times10^{33}},
\end{eqnarray}
and using the alternative aperture correction given in Appendix~\ref{apcor2}
results in
\begin{equation}
\label{o2obssfr2}
{\rm SFR}_{\rm [OII]}\,(M_{\odot}\,{\rm yr^{-1}}) =
  4\pi D_l^{2}\,F_{\rm [OII]}\,10^{-0.4(u_{\rm Petro} - u_{\rm fiber})}
  \left(\frac{S_{\rm H\alpha}/S_{\rm H\beta}}{2.86}\right)^{2.114}
  \frac{1}{2.97\times10^{33}}.
\end{equation}

\noindent The $u$-band luminosity to SFR calibration used is
\begin{equation}
\label{usfrcal}
{\rm SFR_U}\,(M_{\odot}\,{\rm yr^{-1}}) = \left(\frac{L_{\rm U}}
  {1.81\times10^{21}\,{\rm W\,Hz^{-1}}}\right)^{1.186}.
\end{equation}
The derivation given in \S\,\ref{sectionsfru} gives
\begin{equation}
\label{usfr}
{\rm SFR_U}\,(M_{\odot}\,{\rm yr^{-1}}) =
  \left(\frac{10^{-0.4(M_{\rm u}-34.10)}}{1.81\times10^{21}}
   \left(\frac{S_{\rm H\alpha}/S_{\rm H\beta}}{2.86}\right)^{2.061}
  \right)^{1.186}.
\end{equation}
The exponent on the Balmer decrement term here uses the obscuration
curve of \citet{Cal:00}, and incorporates the factor of 0.44 necessary for
obscuration corrections of the stellar continuum \citep[see also][]{Cal:01}.

\noindent For completeness, the SFR calibrations from 1.4\,GHz and FIR
luminosities that give consistent SFR estimates with the above formulae
are also given here \citep[from][]{Bell:03}. The calibration for
1.4\,GHz luminosities is
\begin{equation}
\label{sfr1.4_2}
{\rm SFR_{1.4GHz}}\,(M_{\odot}\,{\rm yr^{-1}}) =
\left\{
\begin{array}{ll}
 L_{\rm 1.4GHz}/[1.81 \times 10^{21}\,({\rm W\,Hz^{-1}})] & L_{\rm 1.4GHz} > L_c \\
 L_{\rm 1.4GHz}/[(0.1 + 0.9 (L_{\rm 1.4GHz}/L_c)^{0.3})\,
   1.81 \times 10^{21}\,({\rm W\,Hz^{-1}})] & L_{\rm 1.4GHz} \le L_c,
\end{array}
\right.
\end{equation}
with $L_c=6.4\times10^{21}\,$W\,Hz$^{-1}$,
and that for FIR luminosities is
\begin{equation}
 \label{belfir2}
{\rm SFR_{FIR}}\,(M_{\odot}\,{\rm yr^{-1}}) =
 L_{\rm FIR}(1 + \sqrt{2.186\times10^{35}\,({\rm W})/L_{\rm FIR}}) /
   [1.85\times10^{36}\,({\rm W})].
\end{equation}

\end{appendix}

\begin{deluxetable}{lcl}
\tablecaption{Scatter of SFR indicators relative to SFR$_{\rm 1.4GHz}$
 \label{sfrtab}}
\tablehead{
\colhead{Indicator} & \colhead{rms scatter} & \colhead{Notes}
}
\startdata
FIR & $\pm40\%$ & Using calibration from Equation~\ref{belfir}\\
H$\alpha$ & $\pm60\%$ & Using obscuration curve from \citet{Car:89}\\
$[$O{\sc ii}$]$ & $\pm70\%$ & Using $F_{\rm OII}/F_{\rm H\alpha}=0.23$\\
$u$-band & $\pm70\%$ & Using derived calibration of Equation~\ref{usfrcal}\\
\enddata
\end{deluxetable}

\clearpage

\clearpage

\begin{figure}
\centerline{\rotatebox{-90}{\includegraphics[width=12cm]{o2rad.ps}}}
\caption{SFR$_{\rm [OII]}$ compared with SFR$_{\rm 1.4GHz}$ and
SFR$_{\rm H\alpha}$. The rms deviation either side of the one-to-one line
is 0.22 dex, a factor of 1.7, for the comparison with SFR$_{\rm 1.4GHz}$,
and is 0.15 dex, a factor of 1.4, for the comparison with SFR$_{\rm H\alpha}$.
Interestingly, the systems with SFR$_{\rm [OII]}\lesssim1\,M_{\odot}$yr$^{-1}$
seem to all show significant overestimates in SFR$_{\rm 1.4GHz}$
(while the SFR$_{\rm H\alpha}$ is consistent with the SFR$_{\rm [OII]}$).
These systems are likely to be hosting a heavily obscured AGN, which dominates
the radio emission but is not detectable through the optical spectroscopic
signature. The error bars in the upper left of both panels indicate 
the typical random error in the measurements.
 \label{o2sfr}}
\end{figure}

\begin{figure}
\centerline{\rotatebox{-90}{\includegraphics[width=12cm]{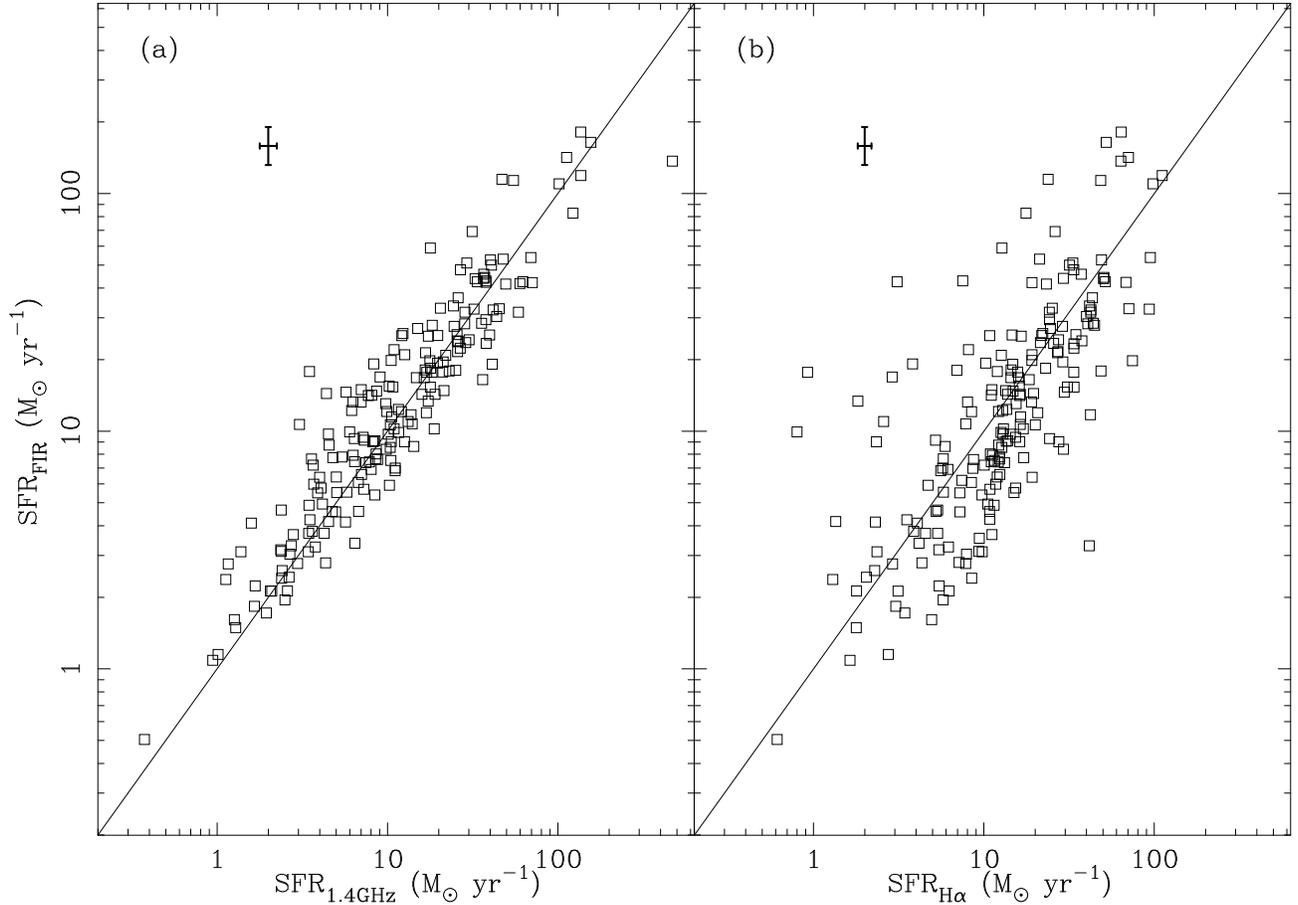}}}
\caption{SFR$_{\rm FIR}$ from Equation~\ref{belfir} compared with
(a) SFR$_{\rm 1.4GHz}$ and (b) SFR$_{\rm H\alpha}$. SFR$_{\rm H\alpha}$ is
calculated as given in Equation~\ref{apobssfr}. The rms deviation either
side of the one-to-one line is 0.15 dex, a factor of 1.4, for the comparison
with SFR$_{\rm 1.4GHz}$, and 0.22 dex, a factor of 1.7 for the comparison with
SFR$_{\rm H\alpha}$. The error bars in the upper left of both panels indicate
the typical random error in the measurements. It is interesting to note
in panel (b) that for SFR$_{\rm FIR}\lesssim7\,M_{\odot}$yr$^{-1}$, almost
all the points show relatively high SFR$_{\rm H\alpha}$/SFR$_{\rm FIR}$.
These galaxies are mostly nearby systems with large aperture corrections
for the H$\alpha$ SFR estimate, and it is likely that the SFR$_{\rm H\alpha}$
is somewhat overestimated for these objects.
 \label{firrad}}
\end{figure}

\begin{figure}
\centerline{\rotatebox{-90}{\includegraphics[width=12cm]{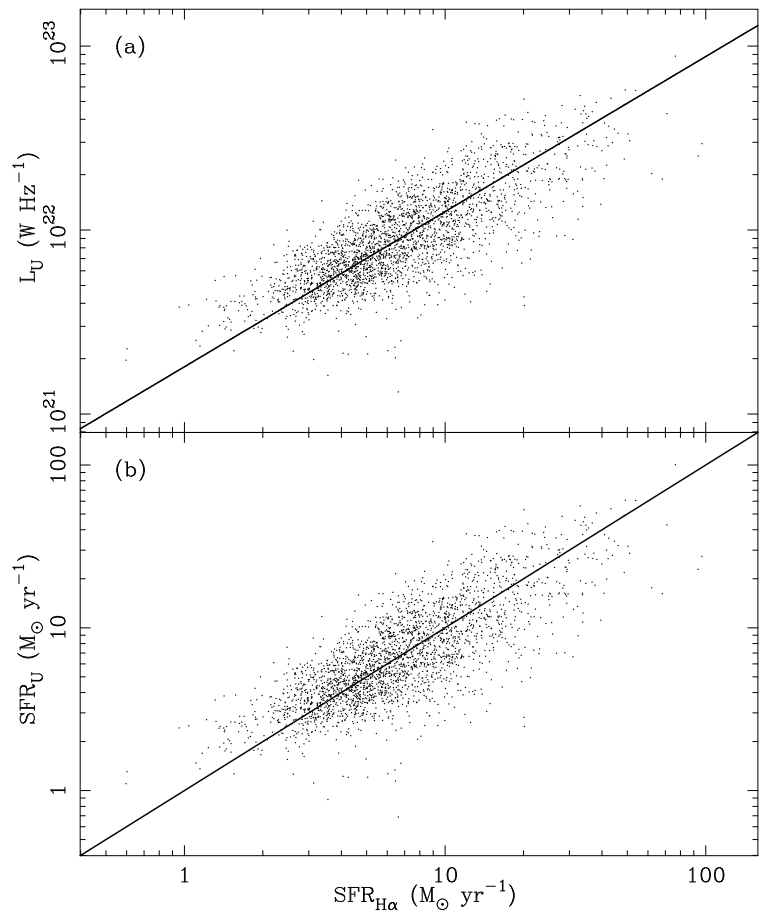}}}
\caption{(a) Comparison between the $u$-band luminosities and the H$\alpha$
SFRs, for the complete sample of 2625 SF galaxies in the DR1.
Luminosities have been corrected for obscuration using the stellar
absorption corrected Balmer decrements. The SFR$_{\rm H\alpha}$ estimates
have also been corrected for aperture effects. The solid line shows the
ordinary least squares bisector fit described in the text.
(b) The resulting SFR comparison after applying the new
SFR$_{\rm U}$ calibration. The rms deviation either side of the one-to-one
line is 0.13 dex, a factor of 1.3. The error bars in the upper left of both
panels indicate the typical random error in the measurements.
 \label{allhausfr}}
\end{figure}

\begin{figure}
\centerline{\rotatebox{-90}{\includegraphics[width=12cm]{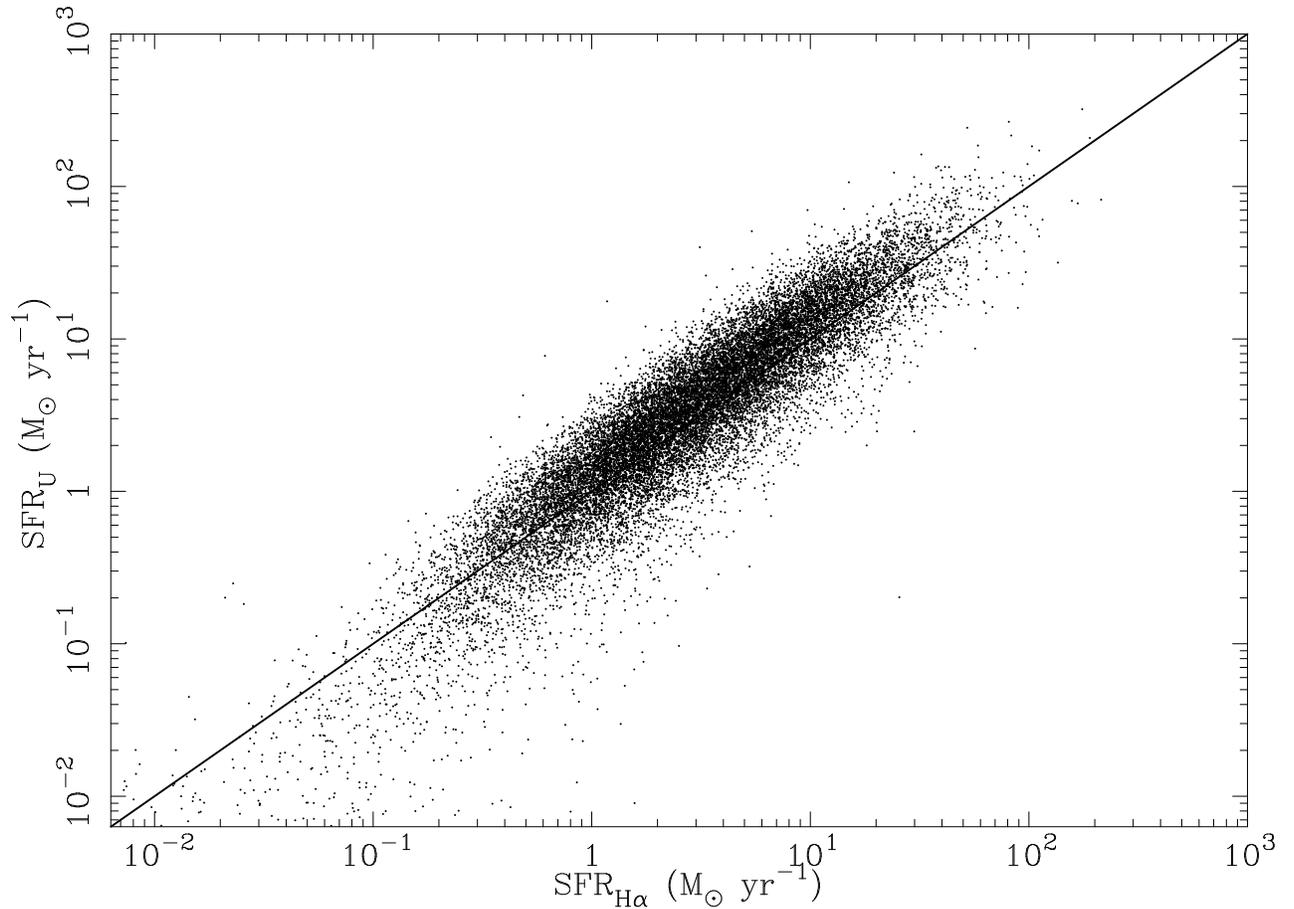}}}
\caption{Comparison between the $u$-band and H$\alpha$ SFRs, for all 21649 SF
galaxies with the necessary measurements in the DR1. Despite the
incompleteness when the entire sample is considered, the derived calibration
is very similar to the ordinary least squares bisector line fit.
There can still be seen, nevertheless, as a slight overall apparent bias
to overestimated SFR$_{\rm U}$ at high SFRs, although the two SFR
estimates are in fact consistent (see Figure~\ref{allhausfr}). 
The error bars in the upper left indicate the typical random error in the
measurements.
 \label{allhausfr2}}
\end{figure}

\begin{figure}
\centerline{\rotatebox{-90}{\includegraphics[width=12cm]{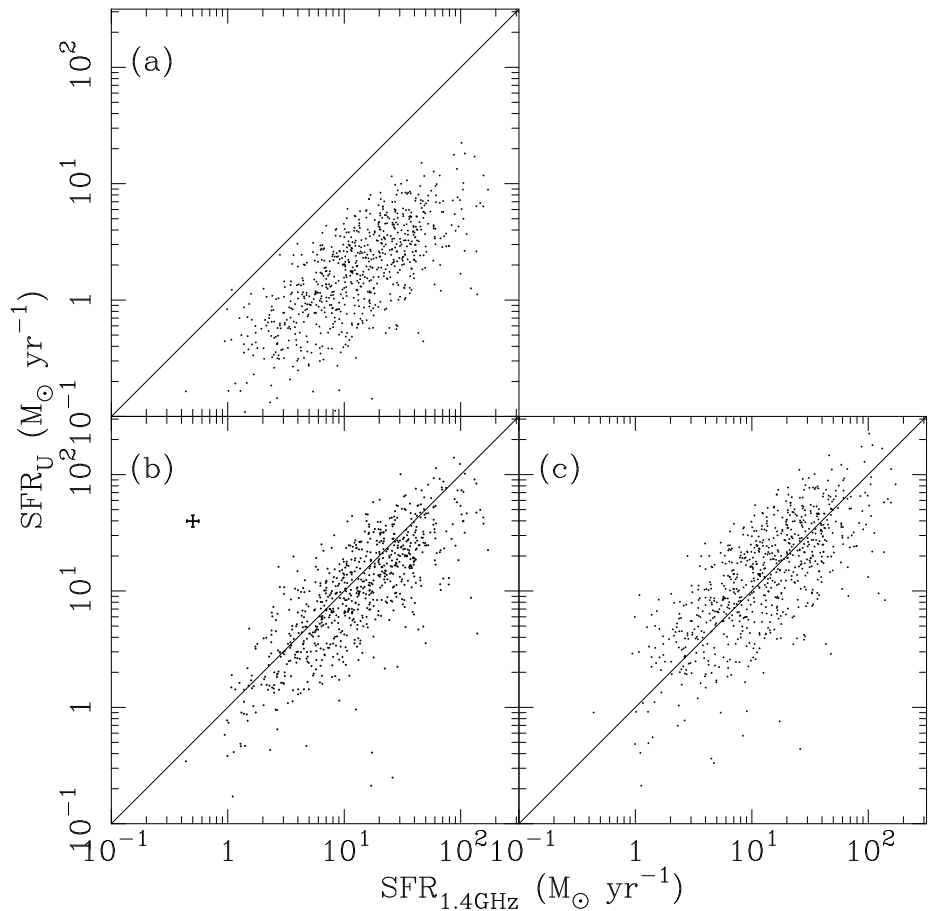}}}
\caption{SFRs from $u$-band luminosity compared with the SFRs from FIRST
1.4\,GHz luminosity. Shown are the 771 SF galaxies remaining after
NVSS measurements are used to replace those from FIRST for galaxies with
$r>10''$. The upper panel (a) shows SFR$_{\rm U}$ before the obscuration
correction is applied. The lower panels incorporate the obscuration correction
to the SFR$_{\rm U}$, directly from the (stellar absorption corrected)
Balmer decrement in (b), and using the method of \citet{Afo:03} in (c).
The error bars in the upper left of panel (b) indicate the typical random
error in the measurements. The slight apparent offset in (b) towards lower
SFR$_{\rm U}$ estimates is a result partially of the incompleteness of this
sample although the same effect is present, to a lesser extent, if the radio
detected galaxies are restricted to those in the complete sample (see
discussion in \S\,\ref{disc_rad}, and Figure~\ref{uvshacplt}).
 \label{firstusfr}}
\end{figure}

\begin{figure}
\centerline{\rotatebox{-90}{\includegraphics[width=12cm]{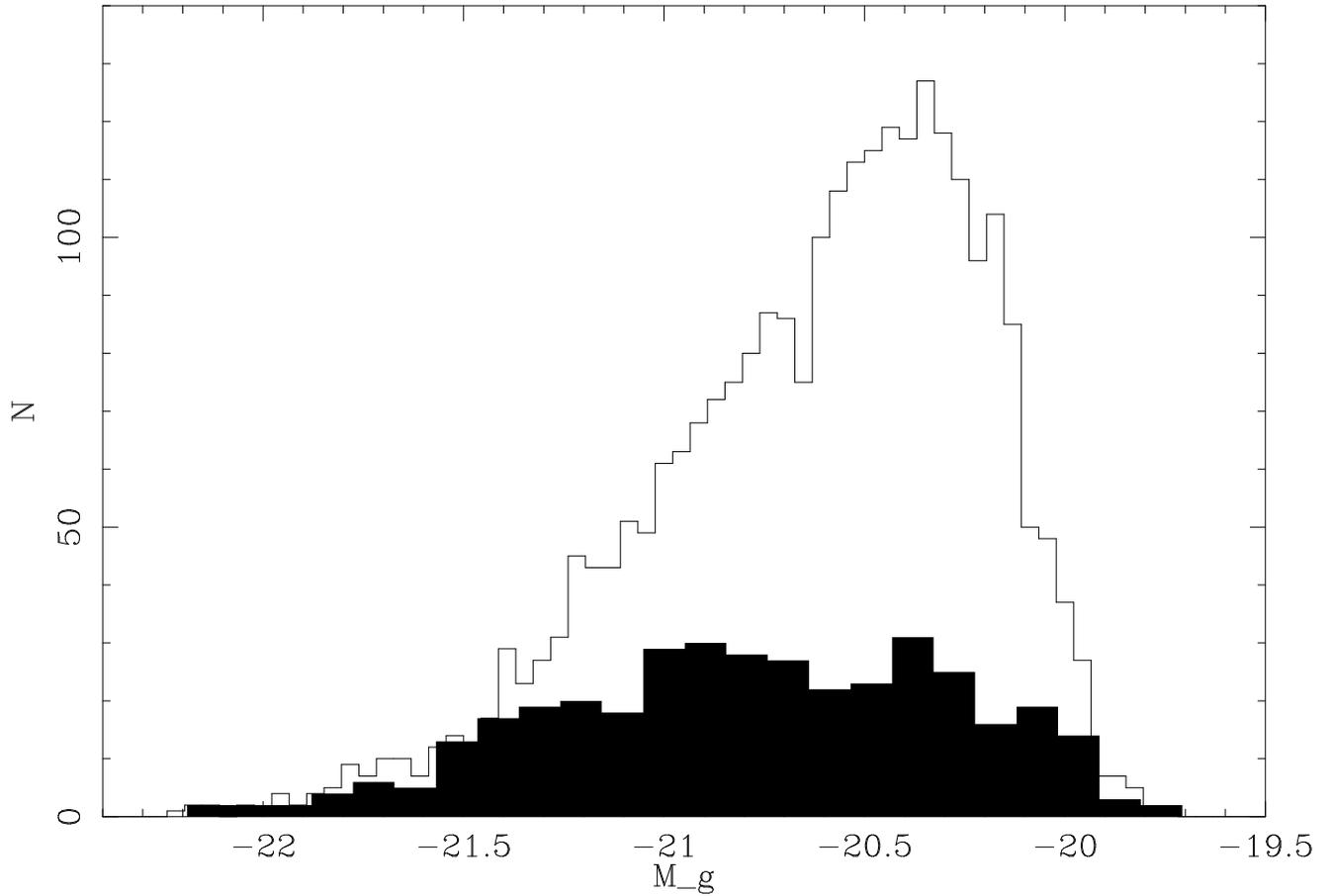}}}
\caption{Histogram showing the range of absolute magnitude, $M_g$, spanned
for the complete sample of 2625 SF galaxies. The radio detected galaxies
within the complete sample are indicated by the filled histogram (note that
different bin sizes have used, to emphasise the shape of the distribution for
the radio detections). For the whole complete sample the median $M_g$ is
$-20.6$, while for the radio detected galaxies it is $-20.8$, although
the distribution for the radio detected systems is much broader. Gaussian
fits to the two distributions here gives a FWHM of 0.9\,mag for the complete
sample, and 1.3\,mag for the radio detected galaxies.
 \label{maghist}}
\end{figure}

\begin{figure}
\centerline{\rotatebox{-90}{\includegraphics[width=12cm]{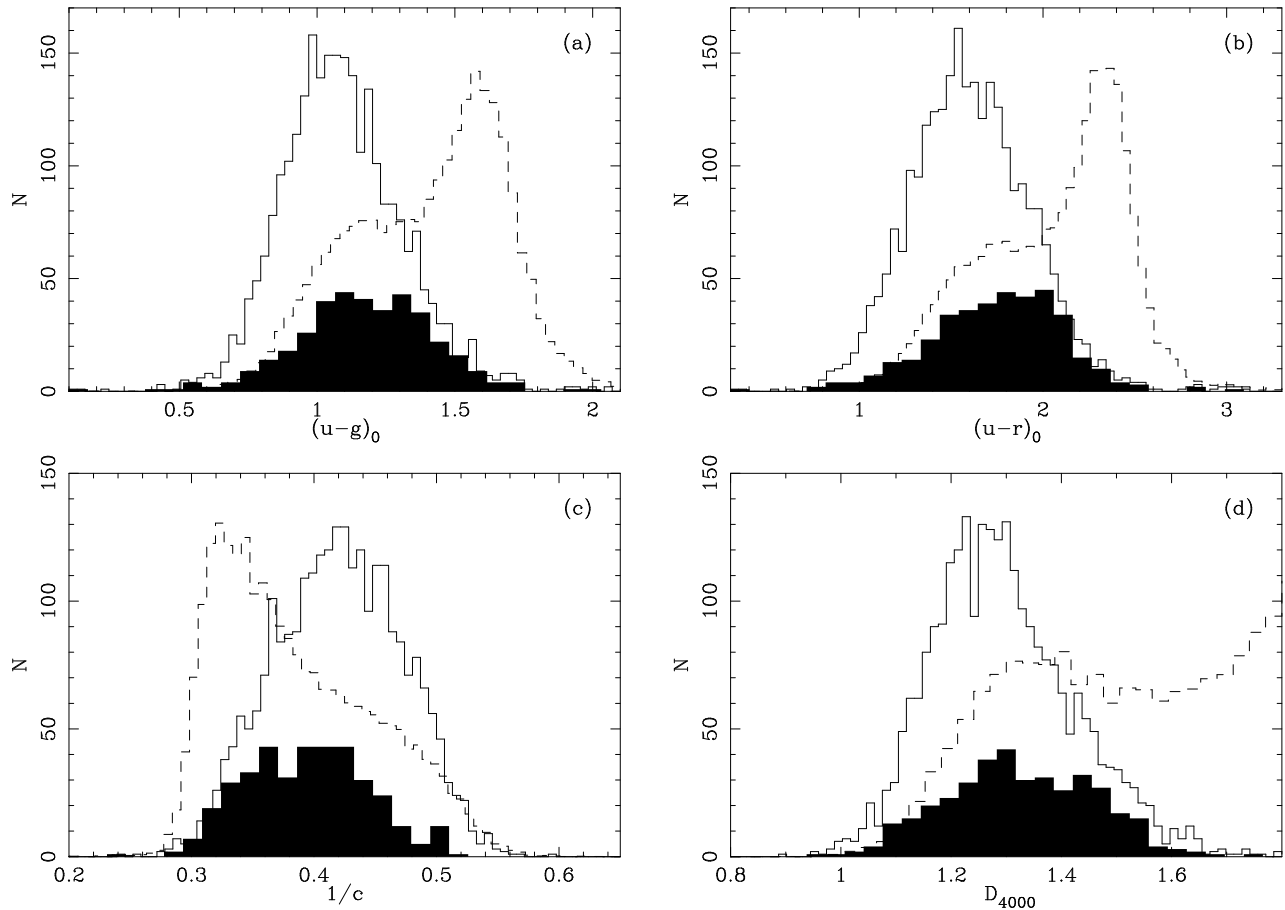}}}
\caption{Histograms showing, for the complete sample of 2625 SF galaxies,
the difference in (a, b) colors, (c) inverse of the concentration index, $1/c$,
and (d) D$_{4000}$. The radio detected galaxies within the complete sample are
indicated by the filled histograms. The dashed histograms in each panel
show the distribution for the complete sample of all galaxies, 24444 objects,
not restricted to star-forming systems. The heights of these histograms have
been scaled down by a factor of ten to emphasise the relative shapes of
the distributions, the bimodal nature of which is clear, being split clearly
into red and blue populations, with the blue population hosting the
star-forming systems. It can be seen that the radio detected
SF population has redder colors, smaller $1/c$ and larger D$_{4000}$ values
than the SF population as a whole. This suggests are larger contribution
to the optical emission, on average, from old-stellar populations in
radio detected SF systems.
 \label{allvsrad}}
\end{figure}

\begin{figure}
\centerline{\rotatebox{-90}{\includegraphics[width=12cm]{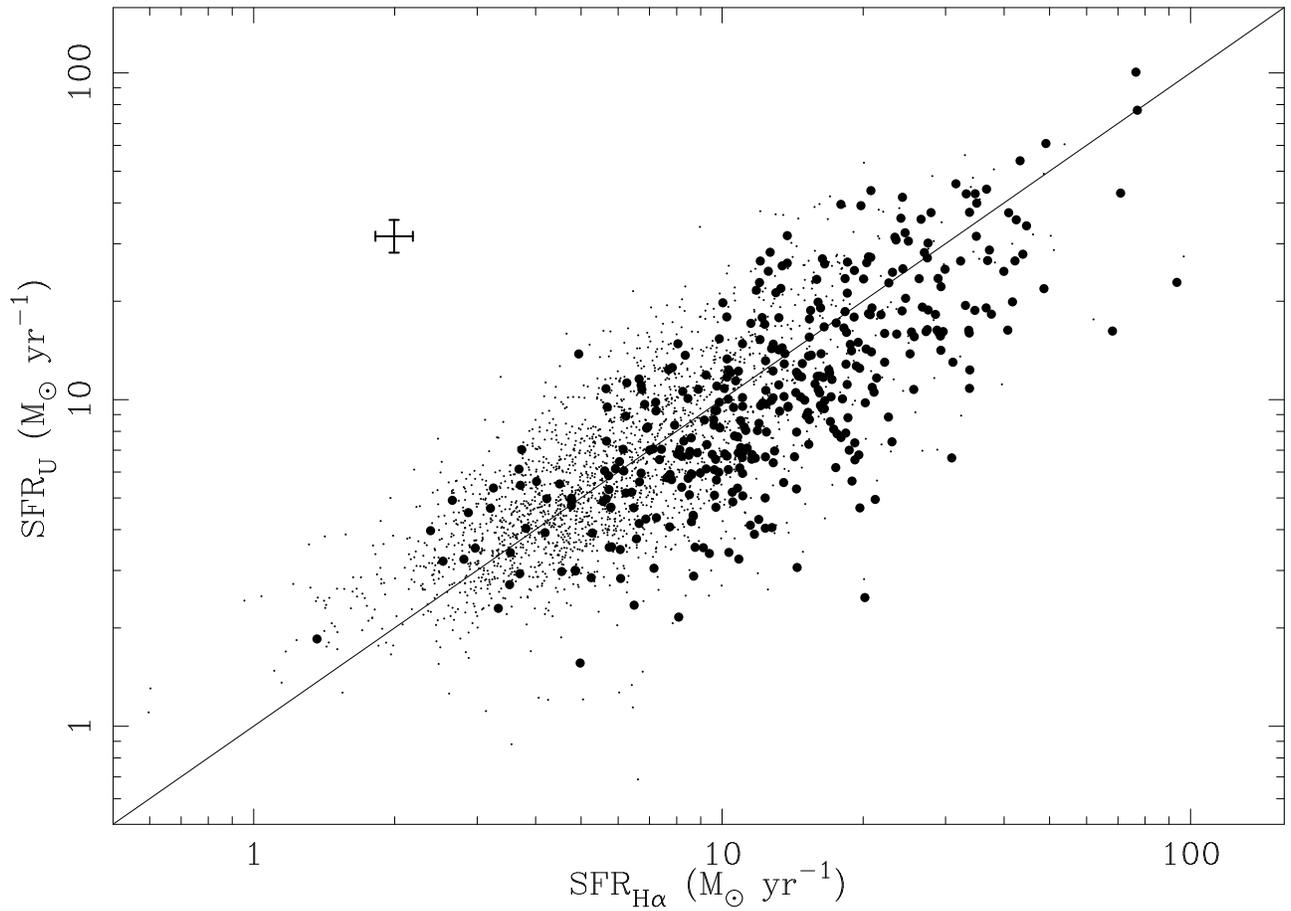}}}
\caption{Comparison between SFR$_{\rm U}$ and SFR$_{\rm H\alpha}$ for
the complete sample (dots), indicating the location of the radio detected
systems (filled circles). The error bars in the upper left indicate the
typical random error in the measurements. It can be seen that, while high SFR
systems ($\approx100\,M_{\odot}$yr$^{-1}$) seem to be uniformly detected at
1.4\,GHz, for moderate SFR galaxies ($\approx10\,M_{\odot}$yr$^{-1}$) radio
detection seems to preferentially select the lower-luminosity $u$-band systems.
 \label{uvshacplt}}
\end{figure}

\begin{figure}
\centerline{\rotatebox{-90}{\includegraphics[width=12cm]{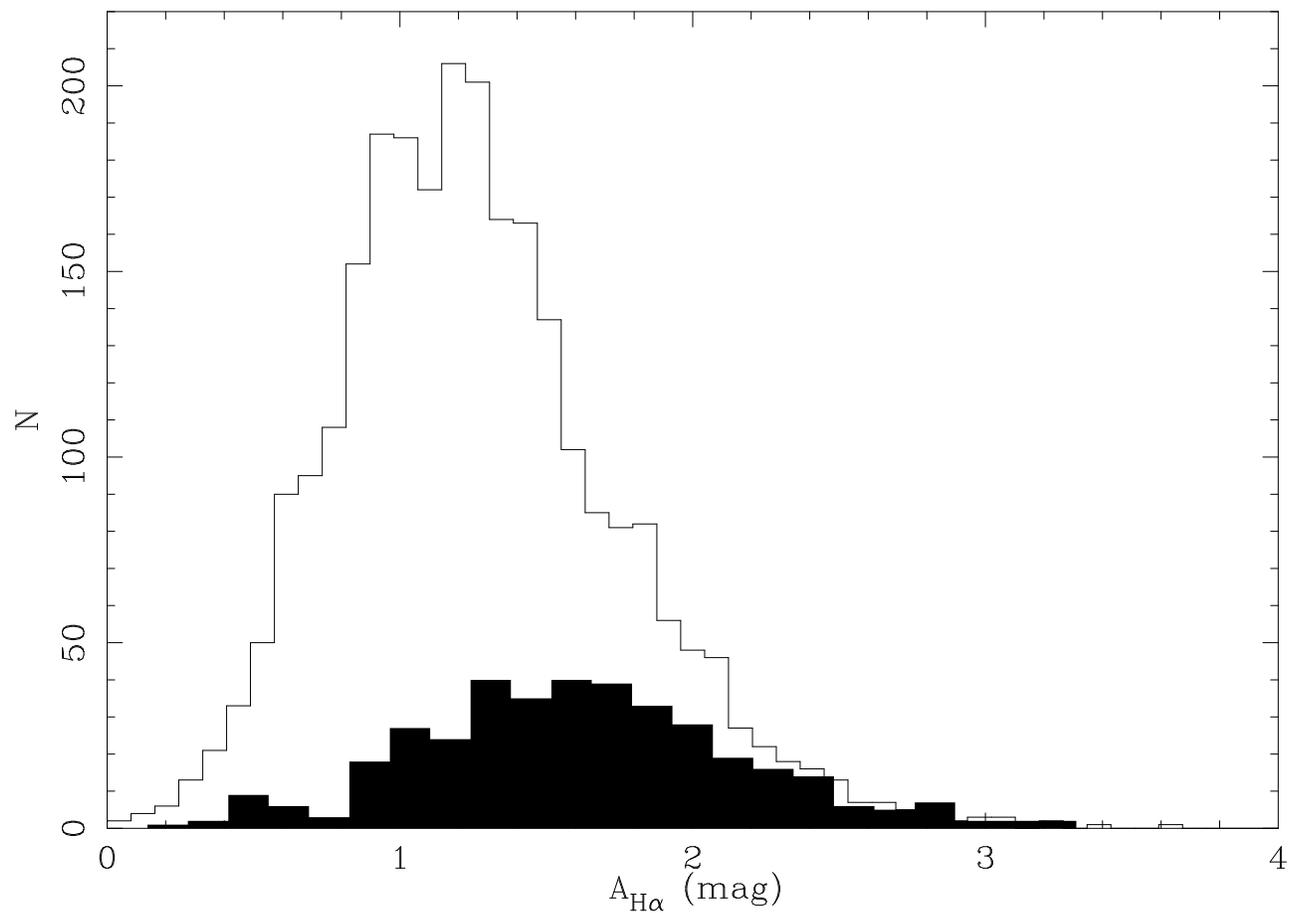}}}
\caption{Histogram showing the (emission line) obscuration in magnitudes
at the wavelength of H$\alpha$ for the complete sample of SF galaxies. The
radio detected galaxies are shown as the filled histogram. The median
obscuration for the complete optically selected sample is 1.2 magnitudes,
while for the radio detected galaxies, it is 1.6 magnitudes. Gaussian fits
to these histograms give FWHM values of 1.0\,mag for the complete
sample, and 1.2\,mag for the radio detected galaxies.
 \label{bdhist}}
\end{figure}

\begin{figure}
\centerline{\rotatebox{-90}{\includegraphics[width=12cm]{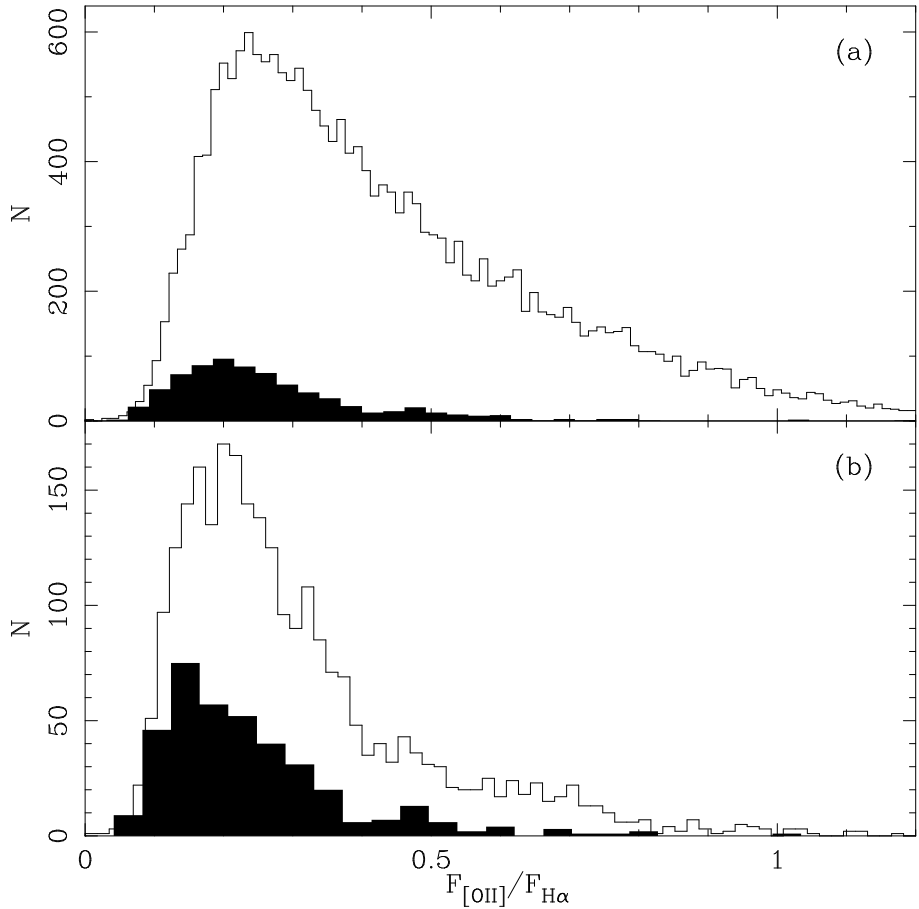}}}
\caption{(a) Histogram of $F_{\rm [OII]}/F_{\rm H\alpha}$ for the full
sample of 22577 SF galaxies from DR1. The 752 radio detected galaxies in the
full sample are shown as the filled histogram. The median value for the whole
(incomplete) sample is $F_{\rm [OII]}/F_{\rm H\alpha}=0.38$, while for
the radio detected sources it is 0.23. (b) The distribution for the complete
sample of SF galaxies. The median value for the complete, optically selected
sample is 0.26. The radio detected galaxies in the complete sample are
shown as the filled histogram, with a median of 0.21.
 \label{o2hist}}
\end{figure}

\begin{figure}
\centerline{\rotatebox{-90}{\includegraphics[width=12cm]{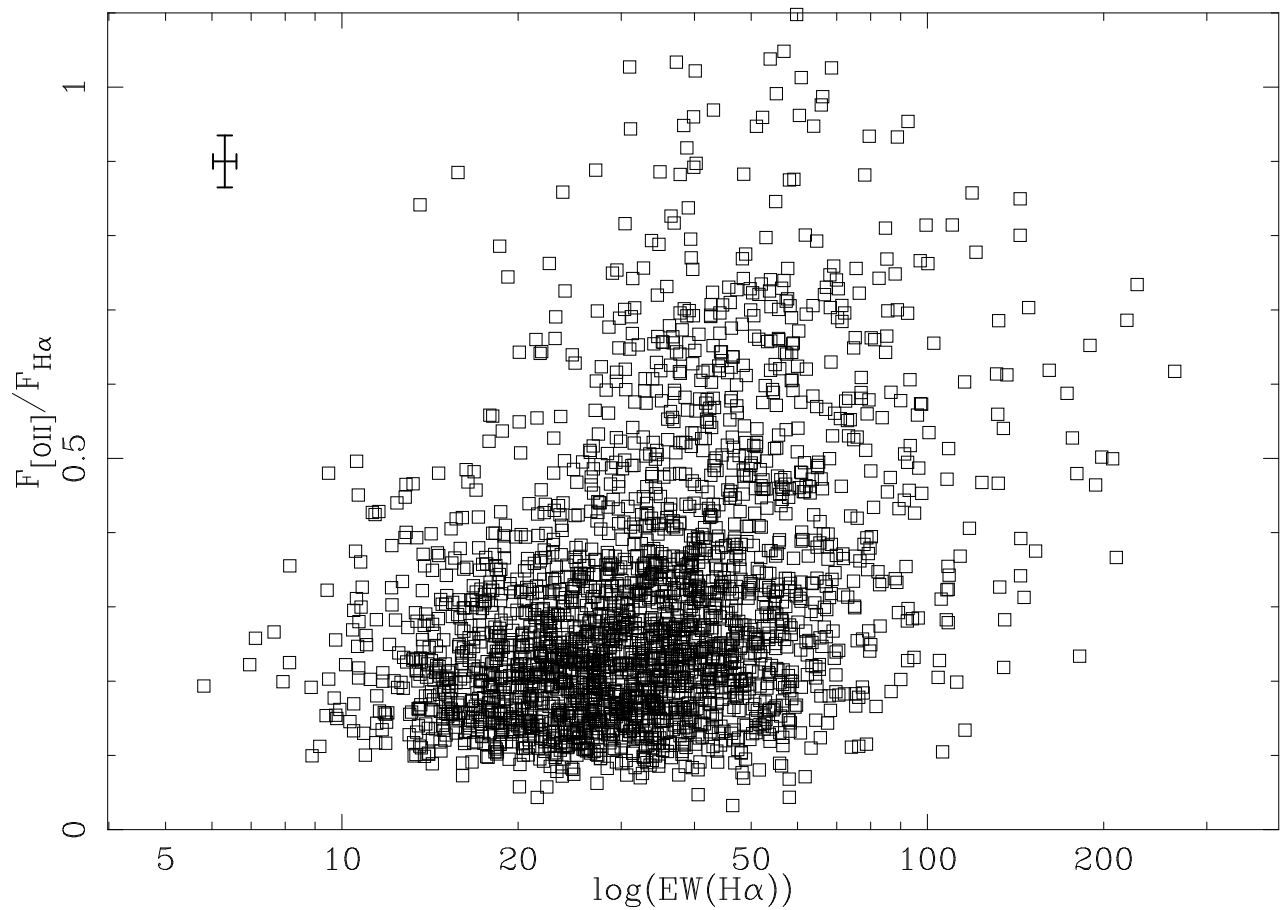}}}
\caption{Ratio of [O{\sc ii}] to H$\alpha$ line fluxes as a function of
EW(H$\alpha$) for the complete sample of SF galaxies. A weak trend can be
seen for higher flux ratios to be present in systems of higher EW.
The error bars in the upper left indicate the typical random error in
the measurements.
 \label{fluxrat2}}
\end{figure}

\begin{figure}
\centerline{\rotatebox{-90}{\includegraphics[width=12cm]{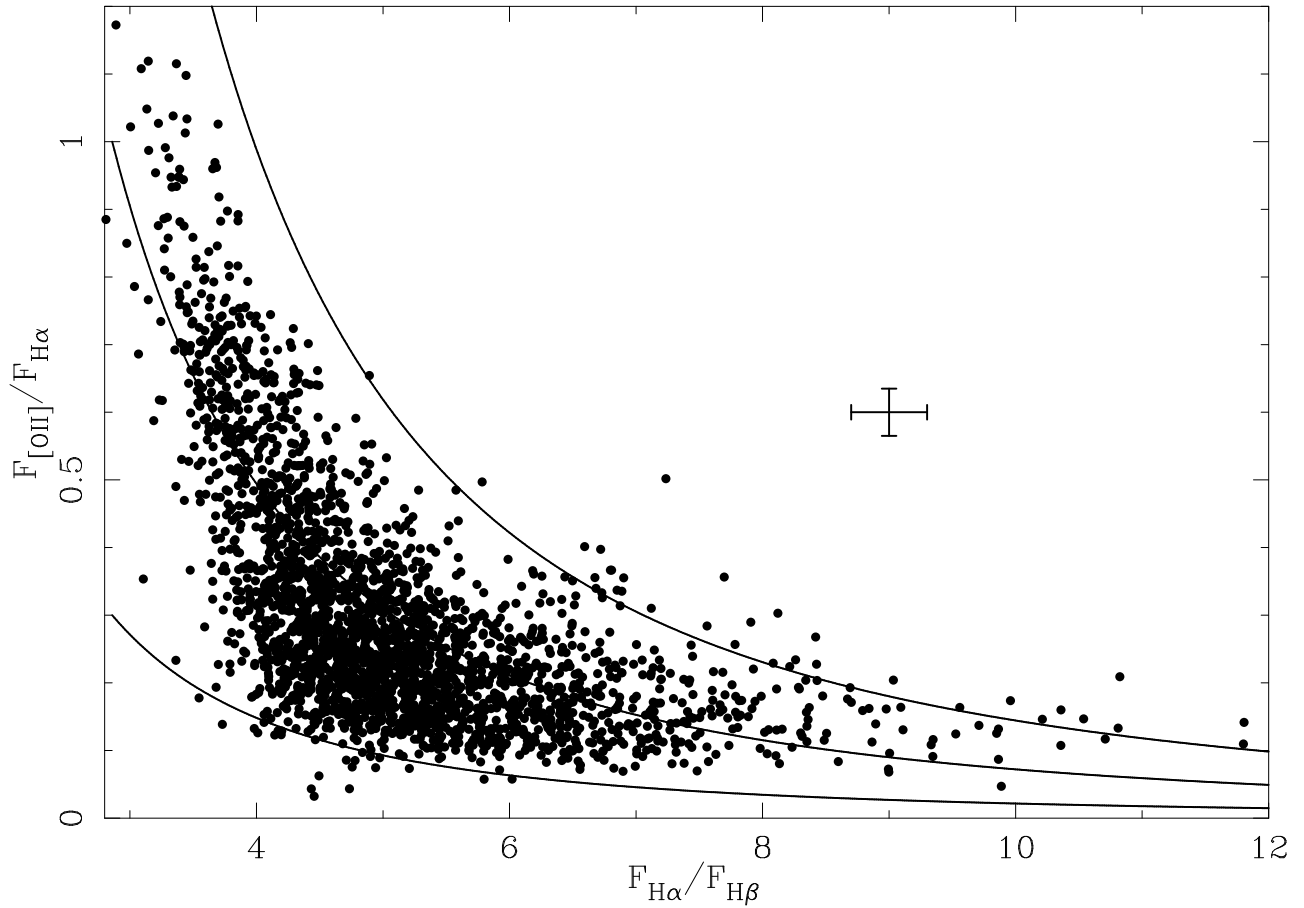}}}
\caption{Relationship between the $F_{\rm [OII]}/F_{\rm H\alpha}$ ratio and
the Balmer decrement for the complete sample of SF galaxies. The trend
seen is to be expected, given that the $F_{\rm [OII]}/F_{\rm H\alpha}$ flux
ratio is constructed prior to obscuration corrections, and hence will trace
the obscuration depending on the obscuration curve used. The error bars in
the upper right indicate the typical random error in the measurements. The
three lines shown in (a) indicate the trend expected from the obscuration curve
of \citet{Car:89} for intrinsic $S_{\rm [OII]}/S_{\rm H\alpha}$ flux ratios
of 0.3, 1.0 and 2.0, from bottom to top respectively.
 \label{o2bd}}
\end{figure}

\begin{figure}
\centerline{\rotatebox{-90}{\includegraphics[width=12cm]{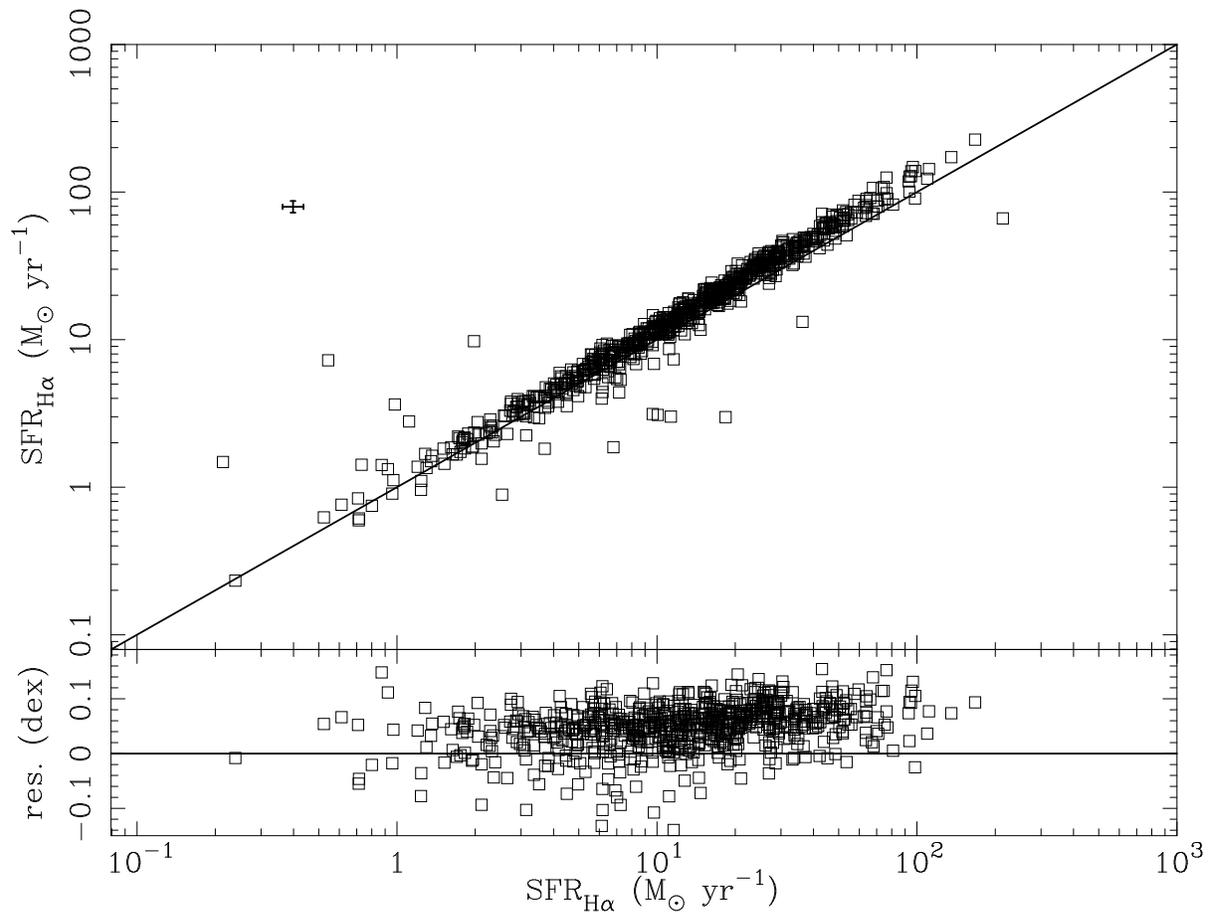}}}
\caption{Comparison of the H$\alpha$ SFRs derived using the different
methods described in Appendix~\ref{apcor2}. The abscissa shows SFR calculated
using Equation~\ref{apobssfr}, based on EW(H$\alpha$) and absolute $r$-band
magnitude, and the ordinate shows SFR calculated using
Equation~\ref{apobssfr2}, based on apparent $r$-band magnitude and fiber
magnitude. Both SFR estimates have been corrected for obscuration using the
Balmer decrement. The error bars in the upper left indicate the typical random
error in the measurements. The lower panel shows the residuals from the
one-to-one line.
 \label{hasfrcomp}}
\end{figure}

\begin{figure}
\centerline{\rotatebox{-90}{\includegraphics[width=12cm]{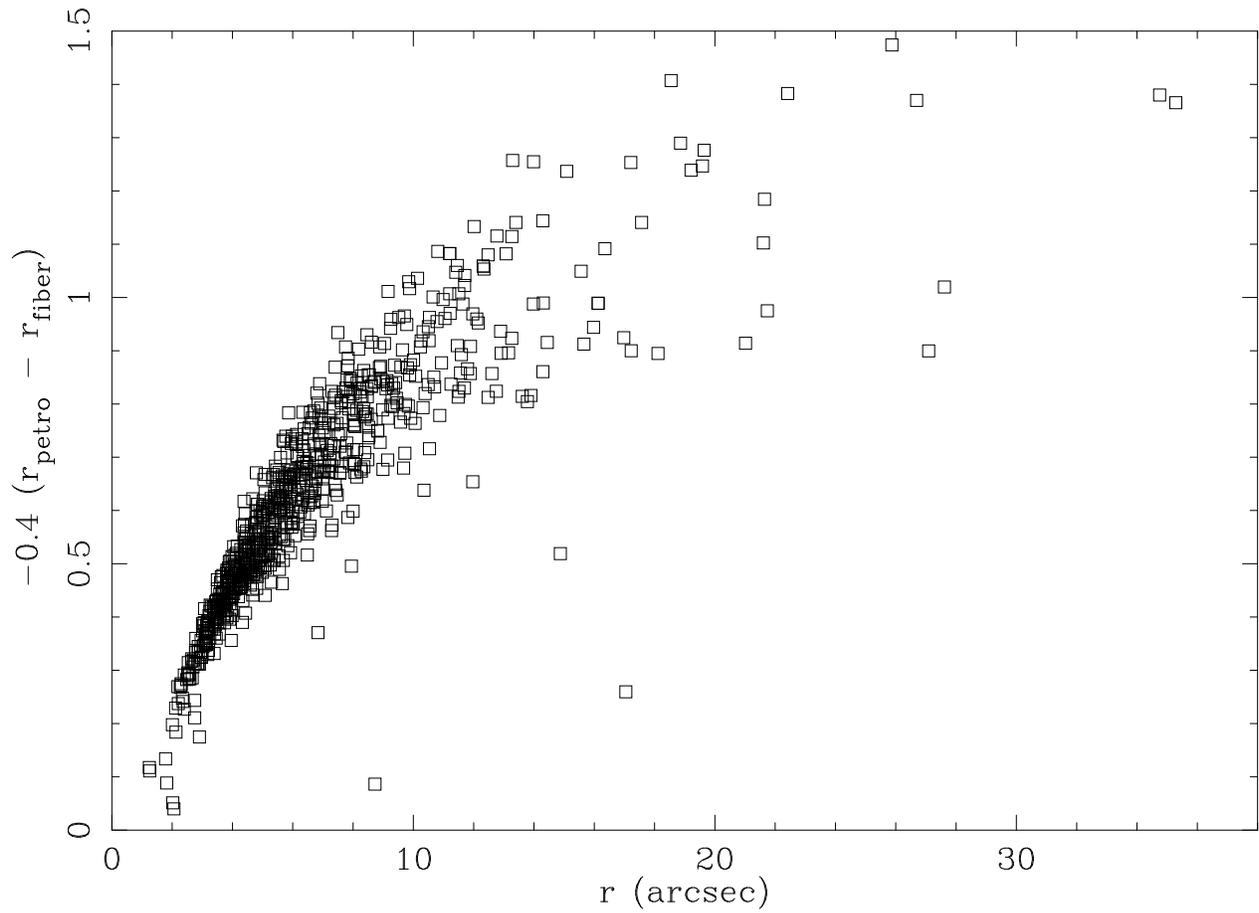}}}
\caption{The aperture correction (from Equation~\ref{apeq}) applied to
the fiber-based SFR estimates shown as a function of galaxy size as given by
the Petrosian radius. Clearly the larger galaxies require a
larger aperture correction.
 \label{apcorr}}
\end{figure}

\begin{figure}
\centerline{\rotatebox{-90}{\includegraphics[width=12cm]{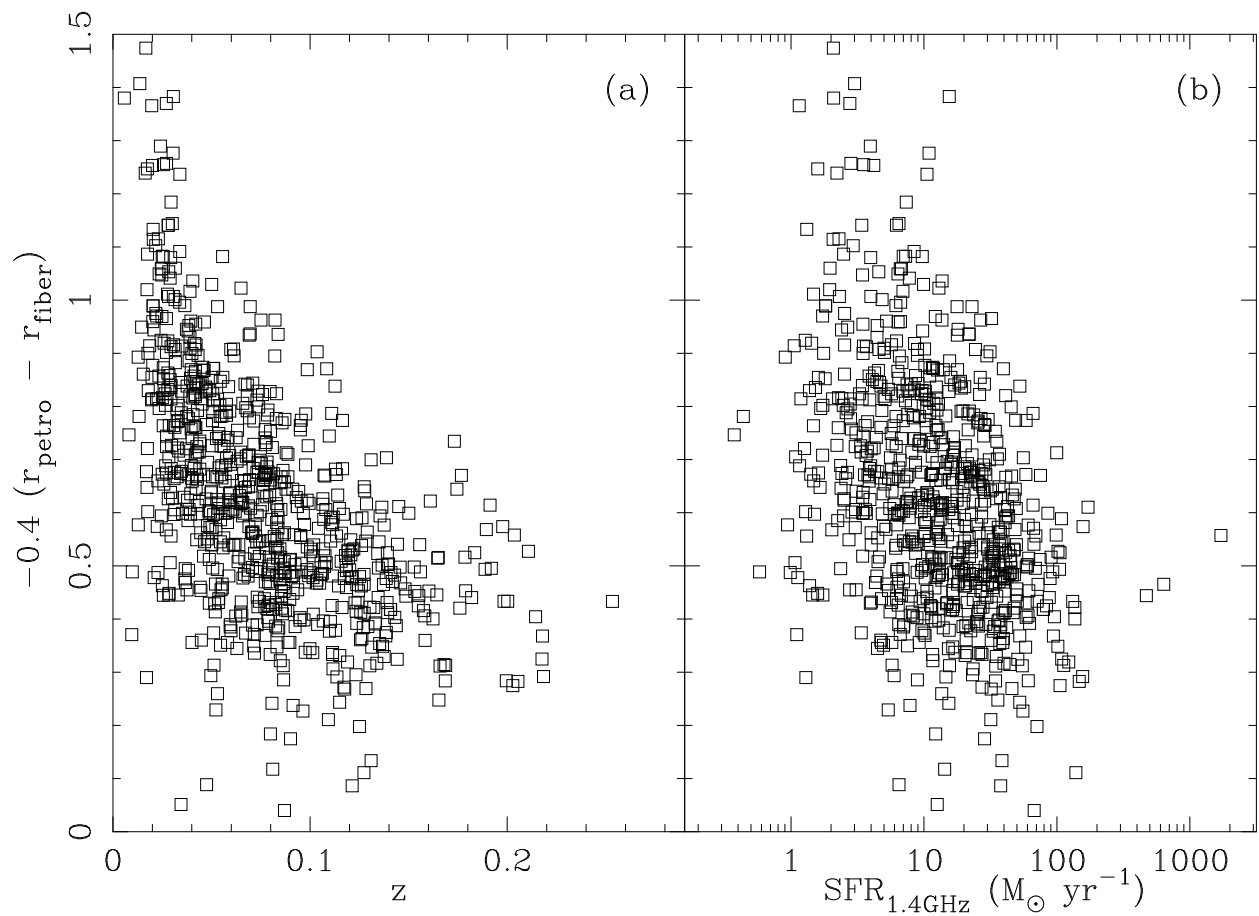}}}
\caption{The aperture correction (from Equation~\ref{apeq}) as a function of
(a) redshift and (b) SFR$_{\rm 1.4GHz}$. The largest galaxies with greatest
aperture corrections lie at the lowest redshifts, and have the lowest
radio derived SFRs. It can be inferred that below about
$10\,M_{\odot}\,$yr$^{-1}$ the FIRST SFRs become progressively further
underestimated. This is shown explicitly in Figure~\ref{sizvssfr}.
 \label{apcorz}}
\end{figure}

\begin{figure}
\centerline{\rotatebox{-90}{\includegraphics[width=12cm]{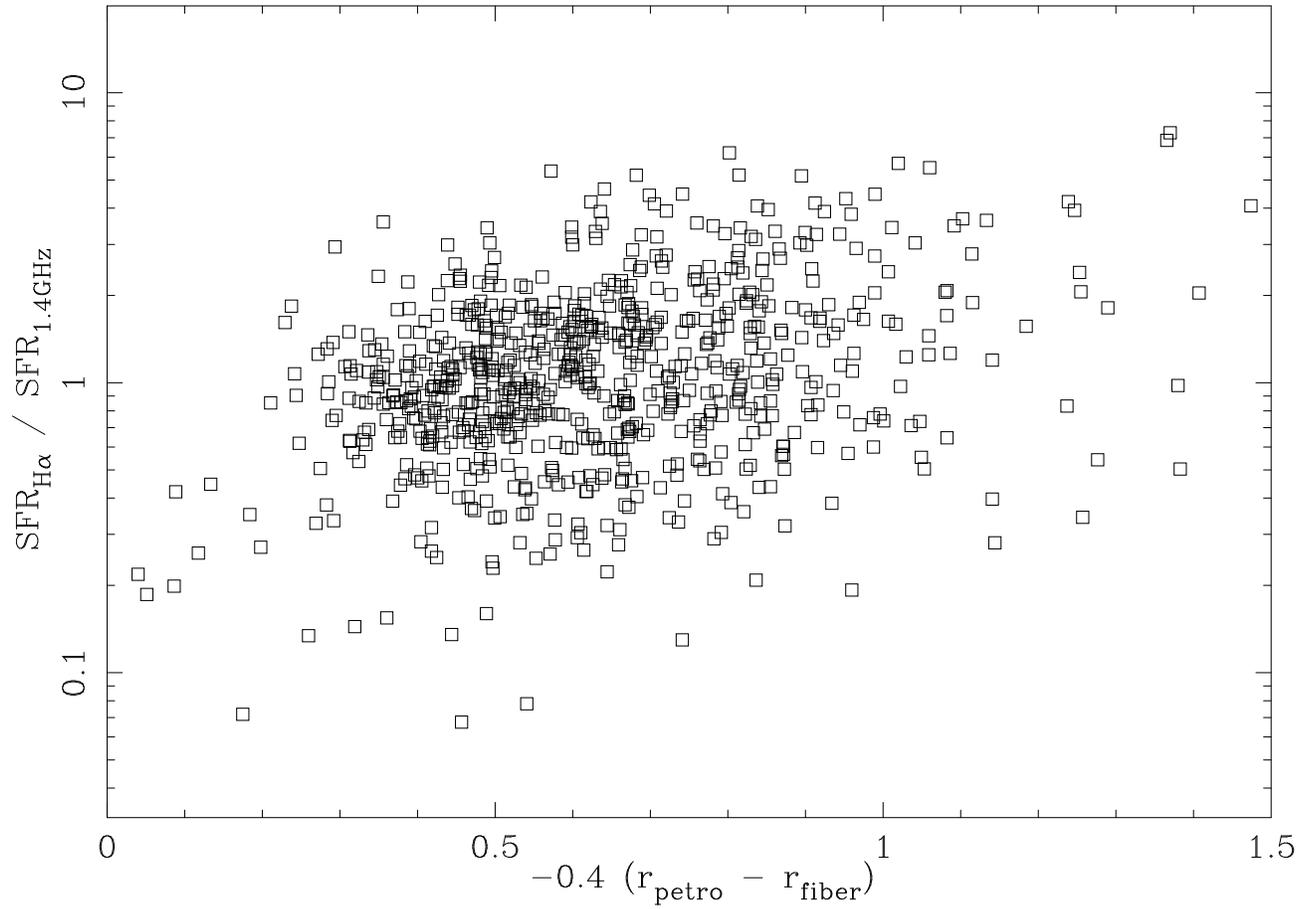}}}
\caption{The ratio of SFRs from H$\alpha$ and 1.4\,GHz luminosities as
a function of the aperture correction (from Equation~\ref{apeq}). The
implicit assumption of a uniform SF distribution made through the
aperture correction results in the slight positive slope seen in this
relation.
 \label{apcorsfrs}}
\end{figure}

\end{document}